%% file: 0_main.tex
\documentclass[journal]{vgtc}                     

\onlineid{0}

\vgtccategory{Research}

\vgtcpapertype{Area 4}

\usepackage{comment}
\graphicspath{{./}{figures/}{./figures/}} 

\usepackage{mathptmx}                  

\usepackage{amsmath}
\usepackage{amssymb}
\usepackage[ruled,linesnumbered]{algorithm2e}

\newcommand{\hfrac}[2]{{\raisebox{.2ex}{$#1$}/\raisebox{-.2ex}{$#2$}}} 
\newcommand{\segment}{\ensuremath{%
  \mathchoice{\includegraphics[height=1.5ex]{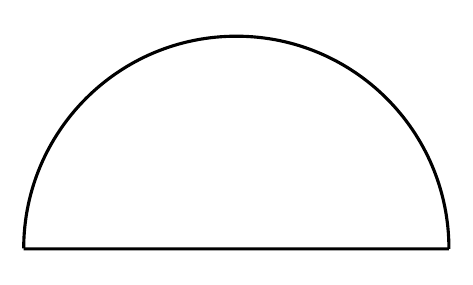}}
    {\includegraphics[height=1.1ex]{segment}}
    {\includegraphics[height=0.8ex]{segment}}
    {\includegraphics[height=0.5ex]{segment}}
}}

\newcommand{\revise}[1]{{\textcolor{black}{#1}}}

\title{Radial Icicle Tree (RIT): Node Separation and Area Constancy}

\author{
    Yuanzhe Jin, 
    Tim J. A. de Jong,
    Martijn Tennekes, and
    \authororcid{Min Chen}{0000-0001-5320-5729}
}
    
\authorfooter{
\item
 Yuanzhe Jin and Min Chen are with the University of Oxford, UK. E-mail: \{yuanzhe.jin, min.chen\}@eng.ox.ac.uk.
\item
 Tim J. A. de Jong and Martijn Tennekes are with Statistics Netherlands, the Netherlands. E-mail: \{tja.dejong,m.tennekes\}@cbs.nl. 
}

\abstract{%
Icicles and sunbursts are two commonly-used visual representations of trees. While icicle trees can map data values faithfully to rectangles of different sizes, often some rectangles are too narrow to be noticed easily. When an icicle tree is transformed into a sunburst tree, the width of each rectangle becomes the length of an annular sector that is usually longer than the original width. While sunburst trees alleviate the problem of narrow rectangles in icicle trees, it no longer maintains the consistency of size encoding. At different tree depths, nodes of the same data values are displayed in annular sections of different sizes in a sunburst tree, though they are represented by rectangles of the same size in an icicle tree. Furthermore, two nodes from different subtrees could sometimes appear as a single node in both icicle trees and sunburst trees.
In this paper, we propose a new visual representation, referred to as \emph{radial icicle tree} (RIT), which transforms the rectangular bounding box of an icicle tree into a circle, circular sector, or annular sector while introducing gaps between nodes and maintaining area constancy for nodes of the same size. We applied the new visual design to several datasets. Both the analytical design process and user-centered evaluation have confirmed that this new design has improved the design of icicles and sunburst trees without introducing any relative demerit.
} 


\teaser{
  \centering
  \includegraphics[width=\linewidth]{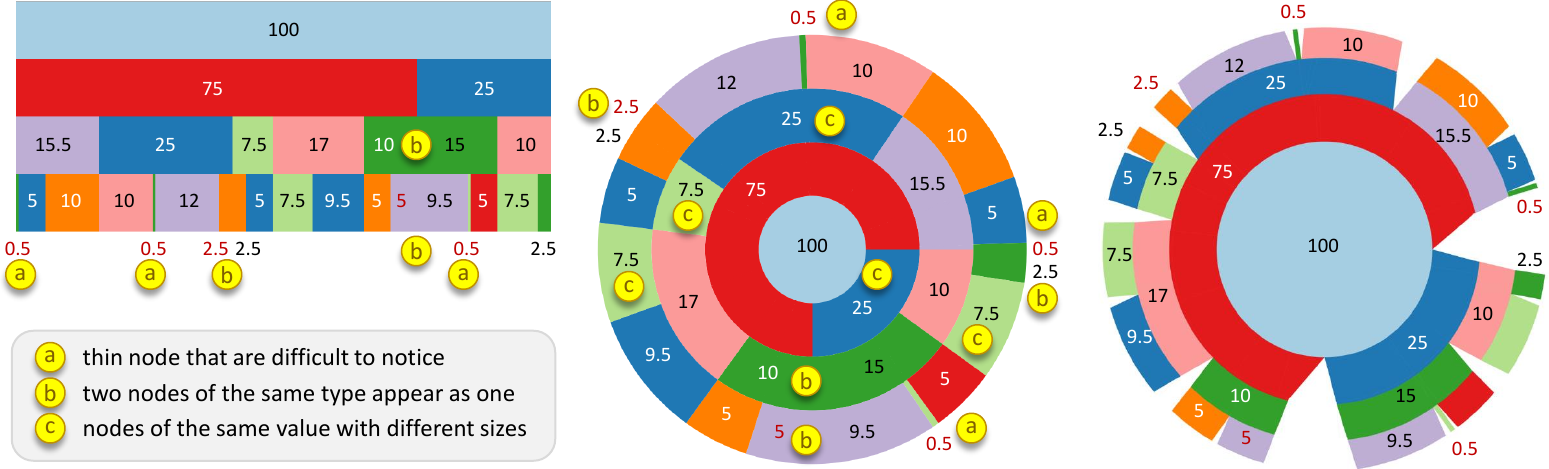}
  \caption{%
  	In a traditional icicle tree plot (left), thin nodes (i.e., nodes of small values) can be difficult to see, and two nodes of the same type (i.e., encoded using the same color) can appear as one node though they belong to different subtrees. A traditional sunburst plot (middle) has a third issue, i.e., nodes of the same value can be mapped to different sizes. These three issues are labeled as (a), (b), (c) in the figure. As shown on the right, a new visual design, referred to as \emph{ Radial Icicle Tree} (RIT) addresses these three issues.%
  }
  \label{fig:Problems}
}

\begin{document}



\firstsection{Introduction\label{sec:Intro}}

\maketitle


\input{1_introduction}
\input{2_related_work}
\input{3_overview}
\input{4_RIT_math}

\input{5_RIT_algo}

\input{6_testing}
\input{7_discussion}
\input{8_conclusion}

\acknowledgments{
This work has been made possible by the Network of European
Data Scientists (NeEDS), a Research and Innovation Staff Exchange
(RISE) project under the Marie Skłodowska-Curie Program.
We would like to express our gratitude to the people who
facilitated this project, in particular, Dolores Romero Morales from
Copenhagen Business School.}

\bibliographystyle{abbrv-doi}

\bibliography{ref}

\newpage
~
\newpage

\appendix

\input{9_AppendixA}
\input{10_AppendixB}

\end{document}

%% file: 1_introduction.tex

Visual clarity and consistency have always been among the desired qualities of data visualization. Many have made strong arguments for maintaining such qualities in most if not all, visual representations. For example, Tufte made a powerful argument of ``graphics integrity'' and coined the term ``lie factor'' to indicate the level at which a visualization image deviated from its source data \cite{Tufte:2001:book}.
Kindlmann and Scheidegger defined three principles, namely ``representation invariance'', ``unambiguous data depiction'', and ``visual-data correspondence'', to formalize the notion of graphical integrity mathematically \cite{Kindlmann:2014:TVCG}. Although such graphics integrity or principles cannot always be maintained by some commonly-used visual representations (e.g., in volume visualization and metro maps \cite{Chen:2022:E1,Chen:2022:E2}), they are nevertheless desired qualities.

\emph{Icicle tree plots} \cite{Kruskal:1983:AS} and
\emph{sunburst tree plots} \cite{Ducheyne:2009:JD}
are two commonly-used visual representations for tree visualization. As illustrated in Fig.~\ref{fig:Problems}, some visual patterns depicted by these plots may exhibit undesirable qualities. As illustrated on the left of Fig.~\ref{fig:Problems}, in an icicle tree, some thin nodes may be difficult to notice (marked as (a) in the figure), and two nodes that belong to different subtrees but are encoded using the same color (e.g., because of same categorical label or semantic type) may visually appear as a single node (marked as (b) in the figure). As illustrated in the middle of Fig.~\ref{fig:Problems}, a sunburst tree may feature nodes that are of the same data values but mapped to visual objects of different sizes (marked as (c) in the figure). Meanwhile, a sunburst tree may also exhibit issues (a) and (b).

For the aforementioned strong arguments of graphics integrity, issue (c) is a serious problem. Even with a more accommodating argument, such as the cost-benefit trade-off \cite{Chen:2016:TVCG}, issue (c) could lead to potential distortion and cognitive cost that should ideally be reduced. Although from the perspective of visual encoding, one could argue that issues (a) and (b) have not breached the aforementioned strong arguments by Tufte \cite{Tufte:2001:book} and Kindlmann and Scheidegger \cite{Kindlmann:2014:TVCG} since the visual objects are encoded with a ``correct'' mapping, and viewers can zoom-in to address the issue (a) and can infer the separation of visually-merged nodes by tracing the separation lines from their parent nodes, grandparent nodes, and so on. For example, in Fig.~\ref{fig:Problems}, presumably, one could infer the separation of the two green nodes (green-10 and green-15) in the sunburst tree (middle-bottom) based on the separation of their parent nodes (red-75 and blue-25). One could then infer the separation of their child nodes pale-purple-5 and pale-purple-9.5. However, with Chen and Golan's argument \cite{Chen:2016:TVCG}, the cost-beneficial ratio would be unfavorable as the effort for cognitive reasoning and interaction would be high.      

In this paper, we propose a new visual representation that bears some resemblance to the designs of icicle trees and sunburst trees but addresses the aforementioned issues mathematically and algorithmically.
In particular, we introduce visual gaps between neighboring nodes to improve the separation of sub-trees and the visibility of thin nodes, while maintaining the numerical consistency in mapping data values to sizes of visual objects (i.e., areas of annuli and annular sectors).
This new visual representation is referred to as \emph{Radial Icicle Tree} (RIT).
We report our design process in Section~\ref{sec:Overview}, where we also present our design rationales by analyzing the symptoms, causes, remedies, and side-effects \cite{Chen:2019:CGF} of major design options considered in the process.
We mathematically confirm the area constancy of our approach in Section~\ref{sec:Math}, and provide a recursive algorithm for drawing an RIT in Section~\ref{sec:Algo}.
In Section~\ref{sec:Testing}, following several images for testing different layouts of RITs, we demonstrate the uses of this new visual representation by applying it to two public-domain datasets as well as utilizing it to study movement data in an application.
In Section~\ref{sec:Discussion} we discuss the limitations of icicle and sunburst trees that RITs could not resolve completely and a few possible variants of RITs that could be studied in the future. This is followed by our concluding remarks in Section~\ref{sec:Conclusions}.

%% file: 2_related_work.tex
\section{Related Work}


\textbf{Tree Visualization} is a subarea of graph/network visualization, but it has engendered a large collection of visual designs, many of which are very different from visual designs for graph and network data \cite{Beck:2017:CGF,Behrisch:2016:CGF,McGee:2019:CGF,Tennekes:2021:CGF}.
A web platform, treevis.net, provides an extensive gallery of tree-structured data visualization methods (TSDV) \cite{Schulz:2011:ICGA}.
The TSDV methods have been classified as implicit, explicit, or hybrid representations, according to how different components of a tree structure are shown \cite{Schulz:2010:TVCG}.
In explicit TSDVs, a tree-structured dataset is represented by a node-link diagram, e.g., \cite{Nguyen:2002:InfoVis,Tan:2007:TVCG,Culy:2010:IV,Rusu:2011:IV,Gou:2011:TVCG,Sadahiro:2014:CEUS,Luo:2011:TVCG,Zellweger:2016:IV,Li:2022:TVCG,Yu:2020:Bioinfo}.

Both \emph{icicle} and \emph{sunburst} trees are implicit TSDV methods because the edges between nodes are not shown explicitly. Schulz et al. provided a comprehensive survey on implicit TSDV methods ~\cite{Schulz:2010:TVCG}.

The development of icicle tree visualization can be traced back to Kruskal and Landwehr \cite{Kruskal:1983:AS} (1983) and Kleiner and Hartigan's castle tree may also be considered as an earlier variant \cite{Kleiner:1981:JASA}.
Its noticeable variants include triangular aggregate treemap \cite{Chuah:1998:InfoVis}, cushioned icicle plot \cite{Chevalier:2007:IWPSE}, and space-reclaiming icicle plots \cite{Wetering:2020:PV}.
In particular, van de Wetering introduced a deformed icicle tree layout to reclaim some space freed by subtrees that do not reach certain depths. This improves the efficiency of space usage, especially at deeper hierarchical levels.

The development of sunburst tree visualization can be traced back to Paul Otlet's depiction of universal decimal classification \cite{Ducheyne:2009:JD} (1901). The hierarchical sector chart by American Society of Mechanical Engineers \cite{ASME:1939:book} (1939), Johnson's polar treemap \cite{Johnson:1993:PhD} (1993), and many subsequent variants may also have influenced the design of sunburst trees, though they do not restrict the placement of each node strictly within a radius range according to the depth of the node. The number of ``cascading'' design variants that introduced such a restriction, including those by Andrews and Heidegger \cite{Andrews:1998:InfoVis}, Chuah \cite{Chuah:1998:InfoVis}, and Stasko and Zhang \cite{Stasko:2000:InfoVis} led to the adoption of sunburst trees as a popular method. Other noticeable variants include disk tree \cite{Chi:1998:CHI}, 3D multivariate sunburst plot \cite{Tekusova:2008:IV}, hyperbolic wheel \cite{Lam:2012:CG}, and sundown chart \cite{Woodburn:2019:VIS}. Yang et al. developed a visualization tool based on sunburst tree visualization for reconfiguration and manipulating hierarchical data \cite{Yang:2002:InfoVis}.

Through empirical studies, Cawthon and Moere, and Muramalla et al.
found that the user performance with icicle tree is better than the sunburst~\cite{cawthon:2007:IV,Muramalla:2017:IADIS}. For multivariate variants of both designs, Zheng and Sadlo found that the sunburst performed better~\cite{Zheng:2021:PacificVis}, presumably due to user acceptance.
Users also found sunburst aesthetically more pleasing~\cite{Muramalla:2017:IADIS,cawthon:2007:IV}.

\vspace{2mm}
\noindent\textbf{Graphics Integrity} is a concept proposed and articulated by Edwards Tufte \cite{Tufte:2001:book}, and it has been widely supported by VIS researchers and practitioners. For example, in IEEE VIS conferences, there have been regular events called ``VisLies'', where participants exhibit visualization imagery or techniques that depict data in misleading or confusing ways.
Kindlmann and Scheidegger presented an algebraic framework to aid the formalization of this concept mathematically and defined three principles for visual encoding, namely ``representation invariance'', ``unambiguous data depiction'', and ``visual-data correspondence'' \cite{Kindlmann:2014:TVCG}.
Correll et al. evidenced the negative impact of ``unfaithful'' visual representations \cite{Correll:2019:TVCG}.

In psychology, a number of empirical studies have been conducted to study the perception of different geometric attributes of shapes (e.g., \cite{Maio:1990:PMS,Morgan:2005:VR,Nachmias:2008:VR,Abbas:2013:VR}).
In visualization, 
Skau and Kosara studied different visual cues used to read pie and donut charts and their impact on the accuracy in perceiving the encoded data values \cite{Skau:2016:CGF,Kosara:2019:VIS}.
Qi and Jing examined the accuracy of length and area perception in shape-based data encoding \cite{Qi:2022:D}. 

Meanwhile, a number of empirical studies have shown that faithfully-encoded visual representations may lead to an incorrect perception of the data being encoded. For example,
Alberts Szafir discovered that the same color could be perceived differently when the host objects change sizes \cite{Szafir:2018:TVCG}.
Schloss et al. discovered that different task performances resulted from colormaps consisting of the same set of colors but mapping concepts to colors differently \cite{Schloss:2019:TVCG}. These suggest that graphics integrity at the encoding stage does not assure correct visualization at the decoding stage. 
 Dasgupta et al. discussed a number of factors (in addition to encoding precision) that may affect the correct interpretation in visualization \cite{Dasgupta:2012:CGF}.
 Kanjanabose et al. showed that for data retrieval tasks, visualization did not outperform data tables in both accuracy and response time \cite{Kanjanabose:2015:CGF}. Wang et al. showed that humans are visually not sensitive to small perturbations of the data \cite{Wang:2020:TVCG}. 
These all led to the question of what is the critical encoding precision for graphics integrity in visualization processes.   

As summarized by Chen and Edwards \cite{Chen:2020:book}, there are two schools of thought in the field of VIS, namely \emph{Isomorphism}, which asserts that ``do not introduce any distortion that is inconsistent with the source data'', and \emph{Polymorphism}, which maintains that ``distortion can be featured in visualization and can bring benefit.'' Chen et al. noticed that the strong argument for graphics integrity (i.e., isomorphism) cannot easily be enforced and some commonly-used visualization techniques (e.g., volume visualization and metro maps) exhibit a fair amount of infringement of graphics integrity \cite{Chen:2022:E1,Chen:2022:E2}, and they reasoned that the viewers' knowledge may help alleviate the potential distortion that may be caused by such unfaithful visual encoding.

Chen and Golan noticed that many-to-one mapping is ubiquitous in visualization as well as in other data intelligence processes (e.g., statistics, algorithms, and human-computer interaction) \cite{Chen:2016:TVCG}. Hence retrieving data values from visualization typically involves a lot of one-to-many mappings, and potential distortion is inevitable. Their cost-benefit ratio suggests a trade-off among \emph{alphabet compression} (a mathematical measure for abstraction, filtering, sorting, etc.), \emph{potential distortion} (a mathematical measure for errors, misinterpretation, biases, etc.), and cost.
Using the cost-benefit analysis, we can ascertain issues (a) and (b) in icicle and sunburst trees as high cost in decoding and issues (c) in sunburst trees as distorted abstraction in encoding and potential distortion in decoding. As a trade-off argument, this is not to say that icicles and sunburst trees should not be used at all but to instigate a question of whether there is a visual representation that offers a better trade-off. In the following sections, we will show that the proposed new visual design RIT can facilitate a positive gain in the trade-off by addressing issues of (a), (b), and (c) without introducing any relative demerit.

%% file: 3_overview.tex
\begin{figure*}[th]
    \centering
    \includegraphics[width=\linewidth]{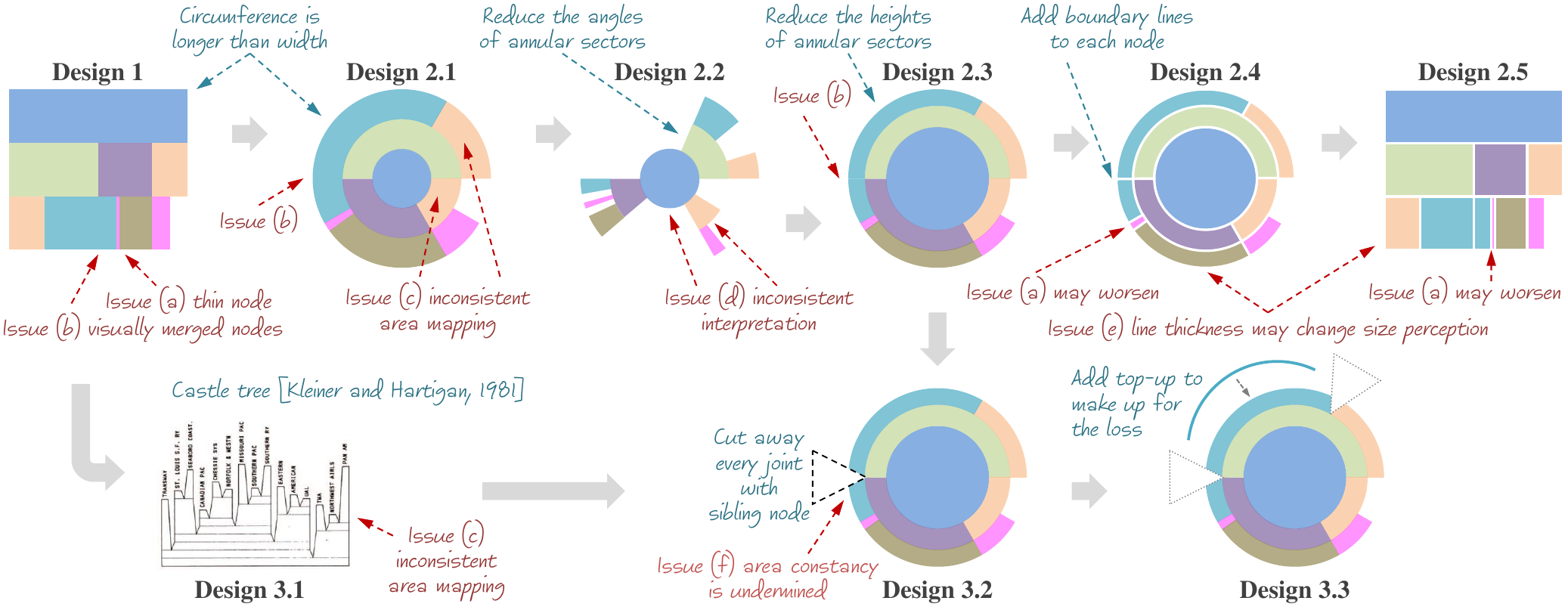}
    \caption{The major designs considered in our design process, where some designs would introduce new issues while addressing the existing issues. Although Design 3.2 was derived from Design 2.3, it is likely that we had unconsciously been influenced by Kleiner and Hartigan's castle tree \cite{Kleiner:1981:JASA}.}
    \label{fig:DesignProcess}
    \vspace{-2mm}
\end{figure*}

\section{Radial Icicle Tree: Design Rationales}
\label{sec:Overview}
As described in Section~\ref{sec:Intro} and as illustrated in Fig.~\ref{fig:Problems}, this work proposes a new visual design for addressing the two issues associated with both icicle and sunburst tree plots and another issue associated only with sunburst trees. In this section, we delineate our process for developing a new design, sketch out the main design options, and describe the design rationales considered in the process. In our design process, we adopted the systematic methodology for improving a visual design (or visual analytics workflow) proposed by Chen and Ebert \cite{Chen:2019:CGF} by using the icicle tree design as the starting point for analyzing symptoms, causes, remedies, and side effects about a visual representation. In Fig.~\ref{fig:DesignProcess}, we illustrate the design process starting from Design 1 on the top-left to Design 3.3 on the bottom-right.

\vspace{2mm}
\noindent\textbf{Design 1:} \emph{Icicle Tree Plot}
\begin{itemize}
    \vspace{-1mm}
    \item \textbf{Symptom:} (a) Hard-to-see thin nodes and (b) visually-merged nodes as illustrated in Fig.~\ref{fig:Problems}. 
    \vspace{-1mm}
    \item \textbf{Cause:} The width of the display space is limited, causing thin nodes and a lack of gaps between subtrees and nodes.
    \vspace{-1mm}
    \item \textbf{Remedy:} Transform it to polar coordinates (i.e., sunburst) to gain more space as the circumference is longer than the width. It can alleviate (a) as some thin nodes become more noticeable. 
    \vspace{-1mm}
    \item \textbf{Side-effect:} It does not address (b) while introducing issue (c) of inconsistent size encoding.
\end{itemize}

We then considered the design of the sunburst tree as \textbf{Design 2.1} in Fig.~\ref{fig:DesignProcess}. The side-effect in \textbf{Design 1} became the symptom in \textbf{Design 2.1}.

\vspace{2mm}
\noindent\textbf{Design 2.1 $\sim$ 2.5} \emph{Sunburst Tree Plot and its Variants}
\begin{itemize}
    \vspace{-1mm}
    \item \textbf{Symptoms:} Aforementioned issues (b) and (c). Issue (a) is not fully addressed.  
    \vspace{-1mm}
    \item \textbf{Cause:} The size encoding based on rectangles is now distorted. The gaps between subtrees and nodes have been abstracted away.
    \vspace{-1mm}
    \item \textbf{Remedy:} One remedy may be to add larger gaps proportionally to ensure area constancy, resulting in Design 2.2 in Fig.~\ref{fig:DesignProcess}.%
    \vspace{-1mm}
    \item \textbf{Side-effect:} A new issue, (d) arises, i.e., the interpretation of these gaps is inconsistent with the conventional interpretation about a portion of the outer edge that is not connected to any child node. Conventionally it would mean that a parent node does pass all of its data value to its child nodes. This interpretation cannot be applied to the ``new type'' of gaps introduced for area constancy.%
    \vspace{-1mm}
    \item \textbf{Remedy:} Instead of reducing the angle of each annular sector, another remedy may be to reduce the height of each annular sector to ensure that the corresponding full annulus has the same area as other full annuli (including the root node that may be depicted as a circle or a full annulus). This is shown as Design 2.3 in Fig.~\ref{fig:DesignProcess}. 
    \vspace{-1mm}
    \item \textbf{Side-effect:} Issue (b) reappears, while the advantage gained by Design 2.1 for addressing the issue (a) starts dissipating because reducing the height of a small annular sector will also make the sector narrower in terms of its arc length.
     \vspace{-1mm}
    \item \textbf{Remedy:} We can add boundary lines between nodes to address (a) and (b) as shown in Design 2.4 in Fig.~\ref{fig:DesignProcess}.
    \vspace{-1mm}
    \item \textbf{Side-effect:} Boundary lines may take up some space that would worsen issue (a) for small and thin nodes. They may change the size perception and such changes will affect small and thin nodes more. We consider this as Issue (e).
    \item \textbf{Remedy:} Boundary lines can also be applied to icicle tree plots as shown in Design 2.5 in Fig.~\ref{fig:DesignProcess}.
    \vspace{-1mm}
    \item \textbf{Side-effect:} Similar to Design 2.4, issue (a) may become worse for small and thin nodes, and issue (e) may occur as the size perception is also affected.
\end{itemize}

In practice, the design variants, such as adding boundary lines to icicles and sunburst trees are relatively common. Nevertheless, the design process led us to a new approach for addressing these issues. The approach is to take the idea of creating bigger gaps from Design 2.2. Instead of removing a portion of an annular sector in the shape of a smaller annular sector (i.e., with a smaller angle), we may remove a fan-like shape such that the inner arc of an annular sector is not shortened, avoiding the creation of issue (d).
Later we discovered that the idea of cutting away a wedge was first reported by Kleiner and Hartigan \cite{Kleiner:1981:JASA} in the context of an icicle tree. Likely, our Design 3.2 might be influenced by their castle tree design, which is shown as Design 3.1 in Fig.~\ref{fig:DesignProcess}.

\vspace{2mm}
\noindent\textbf{Design 3.2 and Design 3.3:} \emph{Radial Icicle Tree}
\begin{itemize}
    \vspace{-1mm}
    \item \textbf{Symptoms:} Design 2.3 addressed issue (c), but not an issue (b). Meanwhile, issue (a) is not fully addressed.
    \vspace{-1mm}
    \item \textbf{Remedy:} Similar to Design 3.1 for an icicle tree, we can also cut away a triangle or fan-like shape at each end of each annular sector. Note that only one of such cuts is illustrated in Design 3.2 in Fig.~\ref{fig:DesignProcess}, though the cuts are to be applied to every annular sector, except a full circle or a full annulus. The cuts enable the separation of nodes and subtrees.   
    \vspace{-1mm}
    \item \textbf{Side-effect:} Naturally, issue (f) arises since the cuts mean size loss, which would undermine our objective of area constancy.
    \vspace{-1mm}
    \item \textbf{Remedy:} Issue (f) can be addressed by adding a thin top-up annular sector to each annular sector that has lost a portion of its area due to the cuts. As long as we can ensure that the top-up sector is of the same size as the lost area, we can maintain area constancy. This is shown as Design 3.3 in Fig.~\ref{fig:DesignProcess}.  
\end{itemize}

After we established that Design 3.3 could address the main symptoms and causes related to issues (a), (b), and (c) as illustrated in Fig.~\ref{fig:Problems} without incurring any significant side-effects as illustrated in Fig.~\ref{fig:DesignProcess}, we needed to consider the mathematical and algorithmic mechanisms for realizing the design. There were still unsolved technical problems, such as how to calculate the lost area that a top-up sector would need to make up for. In Section~\ref{sec:Math}, we will discuss the mathematical properties of Design 3.3, and in Section~\ref{sec:Algo}, we will outline a recursive plotting algorithm for realizing this design.


%% file: 4_RIT_math.tex
\begin{figure}[th]
    \centering
    \begin{tabular}{@{}c@{\hspace{2mm}}c@{}}
        \includegraphics[width=42mm]{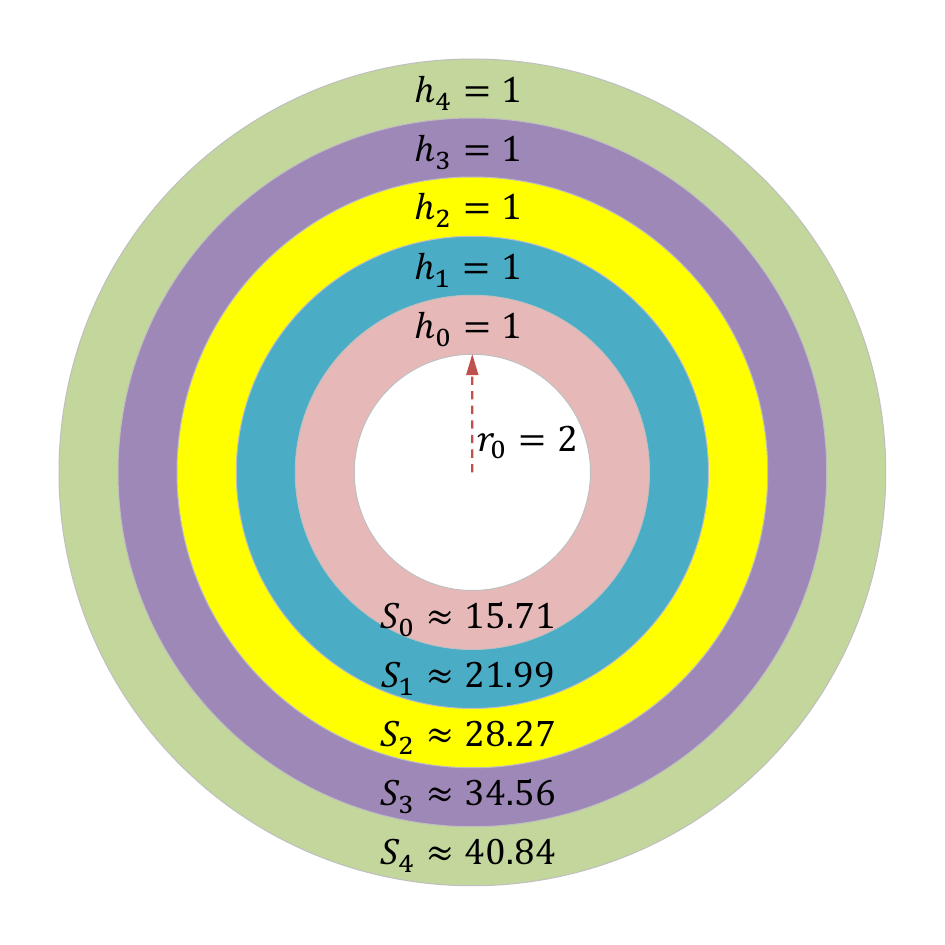} &
        \includegraphics[width=42mm]{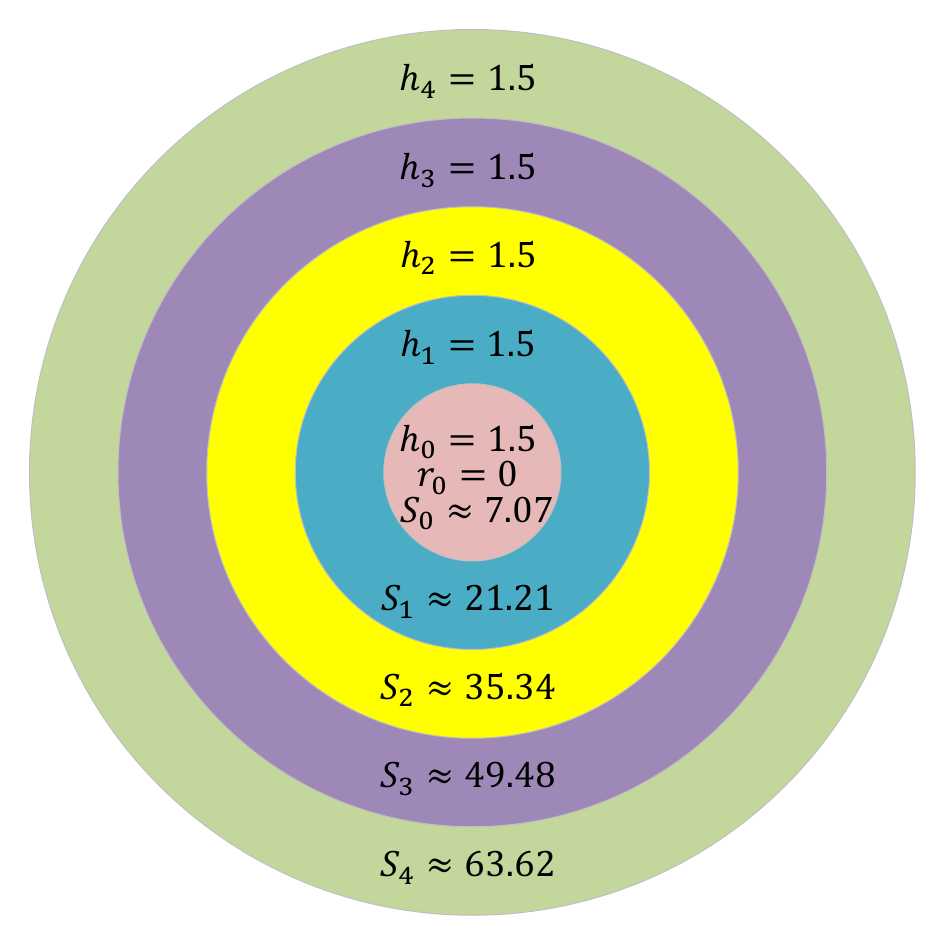} \\
        \small(a) $r_0 = 2; \; h_0 = h_1 = \ldots = 1$ &
        \small(b) $r_0 = 0; \; h_0 = h_1 = \ldots = 1.5$ \\[2mm]
        \includegraphics[width=42mm]{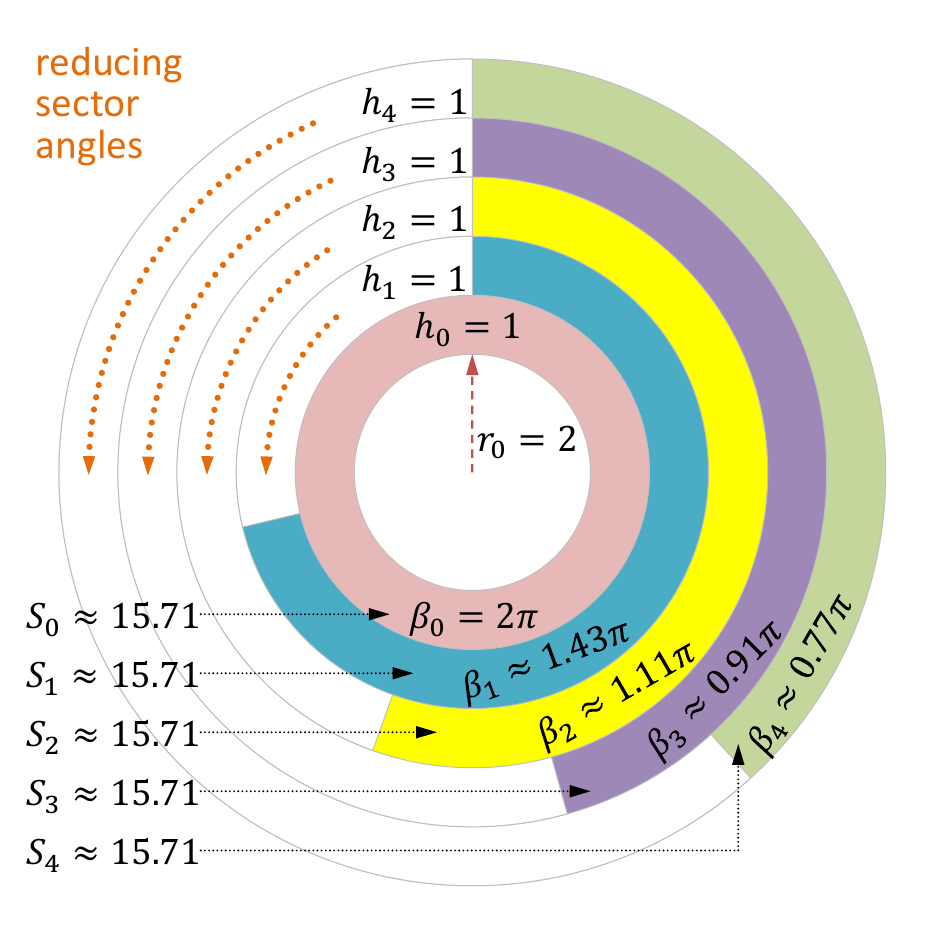} &
        \includegraphics[width=42mm]{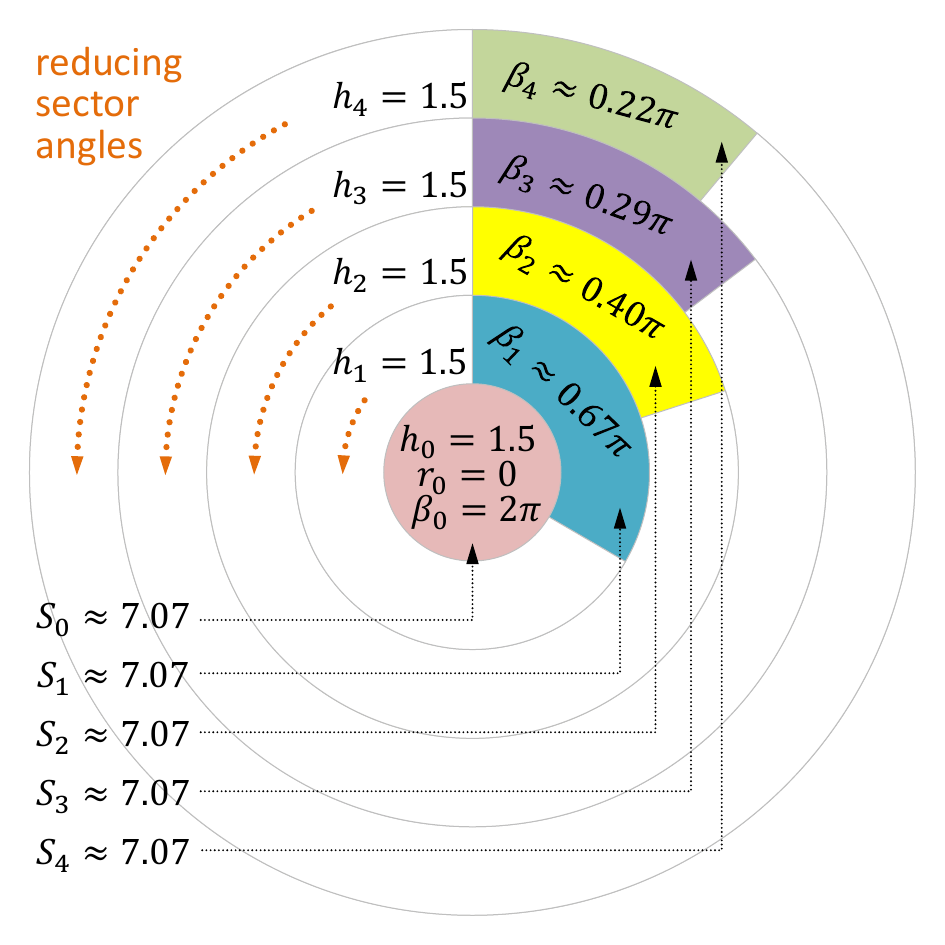} \\
        \small(c) observing size difference in (a)  &
        \small(d) observing size difference in (b) \\[2mm]
        \includegraphics[width=42mm]{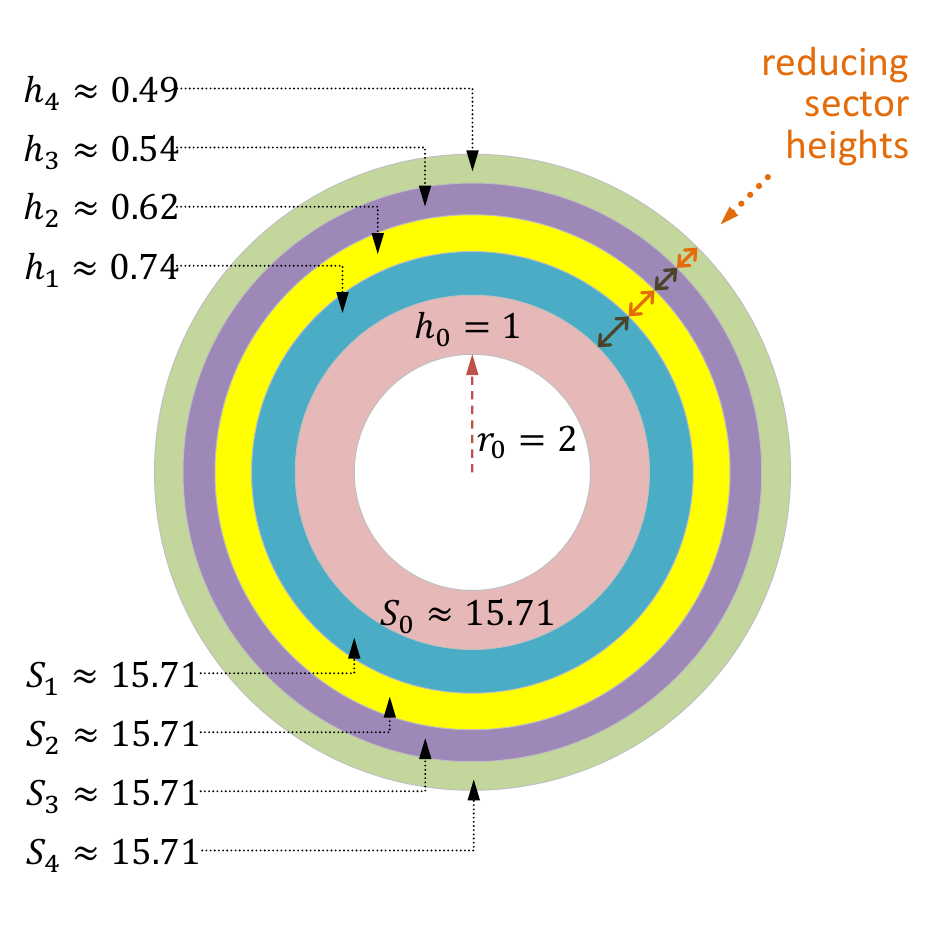} &
        \includegraphics[width=42mm]{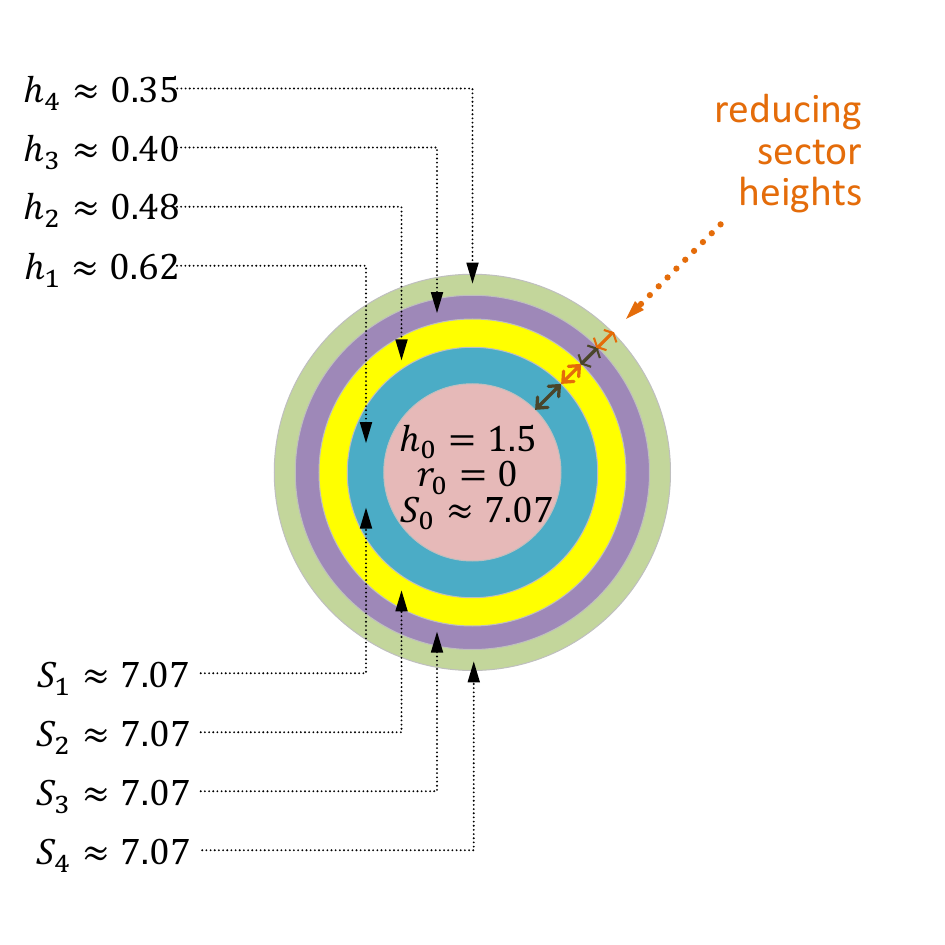} \\
        \small(e) reducing sector heights in (a)  &
        \small(f) reducing sector heights in (b) \\
    \end{tabular}
    \caption{Two examples of size inconsistency are shown in (a, b), where the annuli in each sunburst plot have the same height, causing the increasing areas $S_0 < S_1 < \ldots < S_4$. The scale of inconsistency can be observed more easily in (c, b). One approach to maintain area constancy is to reduce the sector heights gradually, i.e., $h_0 > h_1 > \ldots > h_4$.}
    \label{fig:AreaConstancy}
\end{figure}

\section{Radial Icicle Tree: Mathematical Concepts}
\label{sec:Math}

\subsection{Area Constancy}
\label{sec:AreaConstancy}

The traditional sunburst design does not maintain \emph{area constancy} among different annuli.
Figs. \ref{fig:AreaConstancy}(a, b) show two example sunburst plots depicting two simple trees respectively, each with five nodes.
In each example, each parent node has exactly one child, and all nodes have the same data value of 1 (i.e., 100\%).
In (a), the root annulus (pale red) starts at inner radius $r_0 = 2$, and all nested annuli have the same height, i.e., $h_0 = h_1 = \ldots = h_4 = 1$. The area of each annulus $S_i$ is given at the lower part of the figure, from which we can observe easily that different annuli are of different sizes. In (b), the root annulus is the circle in the middle (i.e., $r_0 = 0$), while $h_0 = h_1 = \ldots = h_4 = 1.5$. Similarly, area constancy is not maintained for this simple sunburst tree.  

One way to observe the scale of the size difference is to maintain the root node annulus as a full annulus in (a) and a full circle in (b), and then change each annulus to an annular sector with an angle $\beta < 2\pi$ such that the area of the annular sector is the same as the area of the corresponding root. 
Figs. \ref{fig:AreaConstancy}(c, d) illustrate such changes. Of course, this will not be a suitable approach for ensuring area constancy because some portion of the outer arc of each node shape (i.e., circle, annulus, or annular sector) is not connected to anything, and it will be interpreted as if a parent node does not pass the full amount of its data value to all of its child nodes, i.e., the data value of the parent is less than the sum of the data values of all  of its child nodes. 

A more suitable approach is to reduce the heights of annular sectors gradually outwards from the root to its leaves in the tree as illustrated in Figs. \ref{fig:AreaConstancy}(e, f). Note that there is only one leaf in each of the two simple examples in the figure.
For each annulus, its area can be computed as:
\begin{equation}\label{eq:AnnulusArea}
    S_i = \pi \bigl( (r_i + h_i)^2 - r_i^2 \bigr), \quad i = 0, 1, \ldots    
\end{equation}
\noindent where $i$ indicates the level of an annulus that corresponds to the depth of a tree such that $i=0$ indicates a tree root and the innermost annulus. Furthermore, $r_{i+1} = r_i + h_i$ with $r_0$ is predefined. In Figs. \ref{fig:AreaConstancy}(a, b), the same annular height was used for all annuli in a plot, i.e., $h_0 = h_1 = \ldots = h_4$, hence causing the increasing annular areas outwards. In Figs. \ref{fig:AreaConstancy}(e, f), $h_0$ is predefined, while $h_1, h_2, \ldots$ are adjusted to ensure all annular areas are the same as $S_0$.

\vspace{2mm}\noindent
\textbf{Theorem 1.}
Let the inner radius of an annulus be a known value $r_i$. When its area is set to a standard size $S_\text{std}$, the height of an annulus is: 
\begin{equation}\label{eq:AnnulusHeight}
    h_i = -r_i + \sqrt{r_i^2 + \hfrac{S_\text{std}}{\pi}}
\end{equation}
\noindent\textbf{Proof.} From Eq.\,\ref{eq:AnnulusArea}, we obtain 
$h_i^2 + 2r_ih_i-\hfrac{S_\text{std}}{\pi}=0$. Because $h_i$ is expected to be $\geq 0$ and since $r_i \geq 0, S_\text{std} \geq 0$, the quadratic equation has only one solution as given in Eq.\,\ref{eq:AnnulusHeight}. $\square$

\vspace{2mm}
With this Theorem, we can derive appropriate values for $h_1, h_2, \ldots$ in Fig.~\ref{fig:AreaConstancy}(e, f), where area $S_0$ is set as the standard size $S_\text{std}$.

Note that the area of an annular sector can be calculated using a formula similar to Eq.\,\ref{eq:AnnulusArea} by replacing $\pi$ with half of the angle of the sector,
i.e., for an annular sector with an arc angle $0 < \beta < 2\pi$, inner radius $r$, and height $h$, its area is: 
\begin{equation}\label{eq:A-SectorArea}
    S_\text{a-sec} = \frac{1}{2} \beta \bigl( (r + h)^2 - r^2 \bigr)
\end{equation}
Given a predefined area $S_\text{a-sec}$ for the sector, we can derive its height using a formula similar to Eq.\,\ref{eq:AnnulusHeight} by replacing $S_\text{std}$ and $\pi$ with $S_\text{a-sec}$ and ${\scriptstyle \frac{1}{2}}\beta$ respectively. In fact, we can use Eq.\,\ref{eq:AnnulusHeight} directly, since:
\[
    \beta S_\text{std} = 2 \pi S_\text{a-sec}
\]

\subsection{Node Separation}
\label{sec:NodeSeparation}
As discussed in Section~\ref{sec:Overview} the icicle design typically packs the nodes more tightly than the sunburst design, often making it hard to distinguish individual nodes. 
In a way, the sunburst design can be seen as a deformed icicle design, where all shapes are collectively transformed from Cartesian Coordinates to polar coordinates. As discussed in Section~\ref{sec:Overview}, this partly alleviates the hard-to-distinguish problem as nodes become wider since the angular axis is usually longer than the horizontal axis. However, unlike a node-link design, there is no space between nodes. Naturally, as illustrated in Fig.~\ref{fig:DesignProcess}, introducing a gap between each pair of neighboring nodes at the same level would address the issue (a) -- the distinguishability of thin nodes, while preventing issue (b) -- the phenomena of merged nodes. 

\begin{figure}[th]
    \centering
    \includegraphics[width=76mm]{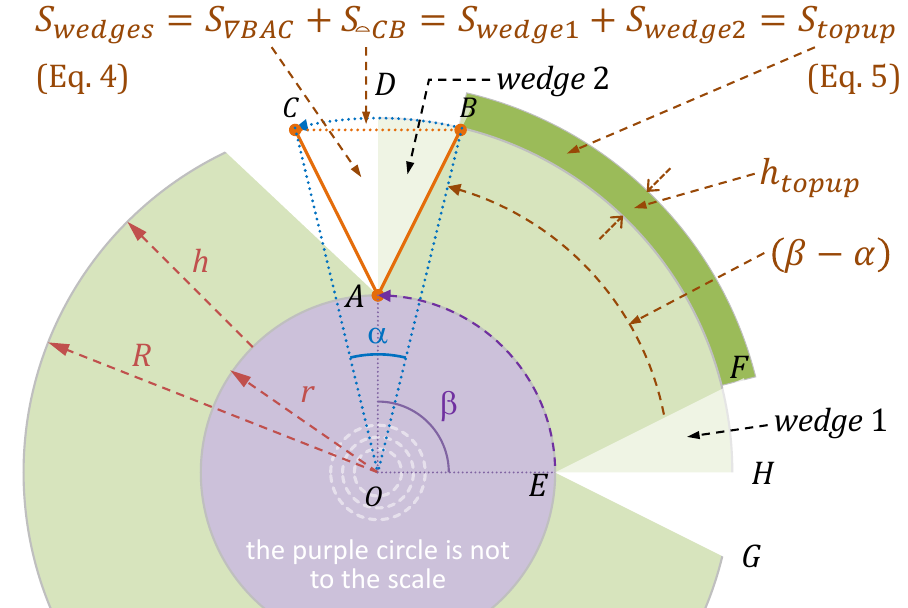}
    \caption{The two wedges to be removed and the top-up annular sector.}
    \label{fig:Wedge}
    \vspace{-4mm}
\end{figure}

In a sunburst design, a node that has a data value of less than 1 (i.e., $<100\%$) is represented by an annular sector. 
One approach is to cut off a wedge at each end of the sector, opening a gap with the neighboring sectors as shown in Fig.~\ref{fig:Wedge}. Let the wedge at one end be mirrored at the other end. For area calculation, we can consider the combined shape that consists of the triangle $\triangledown{BAC}$ and the circular segment $\segment{CB}$ as shown in Fig.~\ref{fig:Wedge}. The area of the two wedges is thus the sum of the area of $\triangledown{BAC}$ and that of $\segment{CB}$. $\segment{CB}, S_{\segment{CB}}$

Let the annular sector be defined with $r$ (inner radius), $h$ (height), and $\beta$ (angle in radian). Let the segment $\segment{CB}$ has an angle $\alpha < \beta$. We can calculate the area of the two wedges to be removed as the sum of the areas of $\triangledown{BAC}$ and $\segment{CB}$, that is:
\begin{equation}\label{eq:WedgeArea}
  \begin{aligned}
     S_\text{wedges} &= S_{\segment{CB}} + S_{\triangledown{BAC}} = S_{\segment{CB}} + (S_{\triangledown{BOC}} - 2S_{\triangledown{BOA}})\\
    &= \frac{1}{2} (r+h)^2(\alpha - \sin \alpha) + \frac{1}{2}(r+h)^2 \sin \alpha - r(r+h) \sin \frac{\alpha}{2}
  \end{aligned}
\end{equation}

Since the removal of the two wedges from the annular sector causes area loss, we need to expand the remaining part of the annular sector somehow to make up for the loss. To maintain the same look-and-feel, we simply add a \emph{top-up} annular sector with the same color and with its area equal to the lost area, i.e., $S_\text{wedges} = S_\text{topup}$. As the outer arc of the remaining part has an angle of $\beta-\alpha$ and its inner radius is $r+h$, we can derive the height of the top-up sector, $h_\text{topup}$, as:
\begin{align}
    S_\text{wedges} =\; &S_\text{topup} = (\beta-\alpha)\bigl( (r+h+h_\text{topup})^2 - (r+h)^2 \bigr) \label{eq:TopupArea}\\
    \Longrightarrow \; &h_\text{topup} = -(r+h) + \sqrt{(r+h)^2 + \hfrac{S_\text{wedges}}{(\beta-\alpha)}} \label{eq:TopupHeight}
\end{align}

\subsection{Maximum of Wedge Angle}
\label{sec:MaxAlpha}

From Fig.~\ref{fig:Wedge}, we can observe that the combined wedge angle $\alpha$ used to remove two wedges at the ends of the annular sector must be smaller than $\beta$.
\revise{If $\alpha > \beta$, the two wedges, $ABD$ and $EFH$ would overlap with each other. If $\alpha = \beta$, the two wedges would join at the middle point along the arc between $B$ and $F$} and it would not be possible to add a top-up annular sector since $\beta-\alpha = 0$ in Eq.\,\ref{eq:TopupHeight}.

In fact, it is not really desirable to have $\alpha$ anywhere near $\beta$ because this would make the top-up sector have a very small angle, which is defined by $\beta-\alpha$. Therefore, we define a wedge angle ratio $ar = \alpha / \beta$. Although $ar$ is mathematically constrained by $0 < ar < 1$, we recommend to maintain a stricter restriction such as $0 < ar < 0.5$. In an implementation, one typically sets $ar$ to 0.1 (i.e., $\alpha$ is 10\% of $\beta$) as a constant for all annular sectors except the root node that does not need wedge removal. One may also decrease $ar$ gradually when the tree depth increases.

\begin{figure}[ht]
    \centering
    \includegraphics[width=85mm]{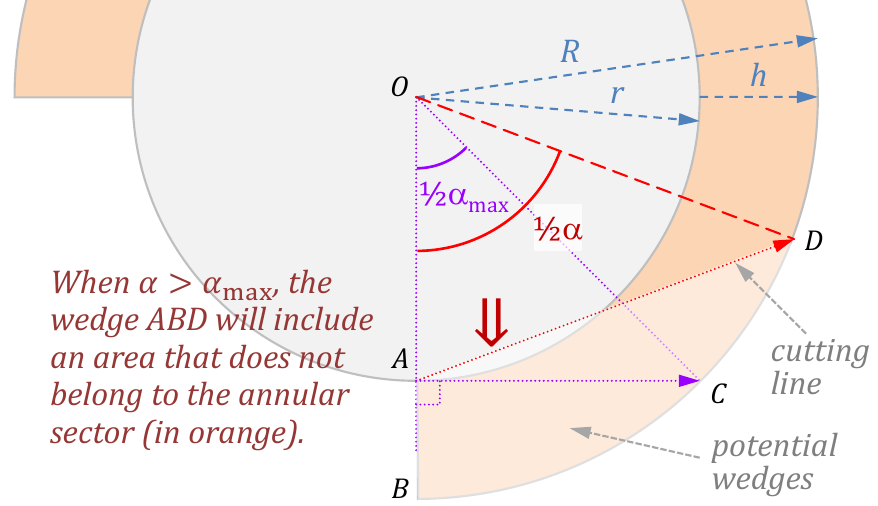}
    \caption{The double-wedge angle $\alpha$ is restricted by two constraints.}
    \label{fig:MaxAlpha}
    \vspace{-4mm}
\end{figure}

Another necessary constraint is illustrated in Fig.~\ref{fig:MaxAlpha}. If one were to remove the wedge $ABD$, the wedge would include an area (indicated by a red arrow \textcolor{red}{$\Downarrow$}) that is not on the annular sector (shown in orange). The constraint is determined by the angle between the lines $AB$ and $AC$, which cannot exceed 90 degrees (or ${\scriptstyle \frac{1}{2}}\pi$ in radian). From the triangle $\triangledown{AOC}$, we can derive:
\[
    \cos(\frac{1}{2}\alpha_\text{max}) = \frac{r}{R} \quad \Longrightarrow \quad \alpha_\text{max} = 2 \cos^{-1} \bigl( \frac{r}{R} \bigr)
\]
We, therefore, recommend the maximum rule for the wedge angle as:
\begin{equation} \label{eq:MaxAlpha}
    0 < \alpha < \min\biggl(\frac{1}{2}\beta, \; 2 \cos^{-1} \bigl( \frac{r}{R} \bigr) \biggr)
\end{equation}

\vspace{2mm}\noindent
\textbf{Theorem 2.}
Given an annular sector defined with $r$ (inner radius), $h$ (height), $\beta$ (sector angle), and $\frac{\alpha}{2}$ (wedge angle), and given that $\alpha$ meets the condition of Eq.\,\ref{eq:MaxAlpha}, we can remove two wedges at the ends of the sector and add a top-up sector defined with $r+h$ (inner radius), $h_\text{topup}$ (height), and $\beta-\alpha$ (sector angle). The resulting shape enables node separation while maintaining area constancy.

\noindent\revise{\textbf{Proof.} In Eq.\,\ref{eq:MaxAlpha}, $\alpha > 0$ assures that two wedges will be removed, hence, assuring node separation. $\alpha < \frac{1}{2}\beta$ assures that a top-up area can be added. $\alpha < \alpha_\text{max}$ assures that the area lost by an annular sector is exactly the area of the two wedges $S_\text{wedges}$.}  

\revise{Theorem 1 assures area constancy before removing any wedge. Eqs.\,\ref{eq:WedgeArea}, \ref{eq:TopupArea}, and \ref{eq:TopupHeight} assure that the area of the two removed wedges $S_\text{wedges}$ is the same as the area of the top-up sector $S_\text{topup}$. Hence the area constancy is maintained after removing the two wedges and replacing them with a top-up sector according to Eqs.\,\ref{eq:WedgeArea}, \ref{eq:TopupArea}, and \ref{eq:TopupHeight}. $\square$}

%


%% file: 5_RIT_algo.tex
\section{Radial Icicle Tree: Recursive Drawing Algorithm}
\label{sec:Algo}
Tree drawing algorithms often make use of a recursive procedure for drawing subtrees. We adopt the same approach in the design of an algorithm for drawing radial icicle trees (RITs). The algorithm starts with the main procedure \texttt{draw\_rit()} that draws the root node as an annulus or an annular sector. It then recursively invokes the procedure \texttt{draw\_subtree()} for drawing its child nodes. After drawing each child node, \texttt{draw\_subtree()} recursively invokes itself for drawing the child nodes of the node that has just been drawn.

The algorithm assumes that the tree data structure consists of a list of nodes, and the data of each node is accessed through a handler $n$, which can be an index or a pointer depending on the implementation. The data record of each node contains several fields, \revise{including
$n.Children$ (handlers of child nodes), $n.Data$ (data value), $n.Color$ (node color), $n.\theta$ (starting angle of an annular sector), $n.\beta$ (arc angle of an annular sector), and $n.\alpha$ (double-wedge angle of an annular sector). A more detailed description of these fields can be found in Appendix~\ref{app:NodeRecord}.}

\begin{algorithm}
\caption{The main RIT algorithm: \texttt{draw\_rit()}}\label{alg:Main}
  \KwData{$n, \; \theta_0, \; \beta_0, \; \; r_0, \; h_0, \; C_{ar0}, \; C_\text{acr}$}
  draw\_a\_sector($n.\text{Color}, \; \theta_0, \; \beta_0 \; r_0, \; h_0$)\;
  $A_\text{std} \gets 0.5 \cdot \beta_0 \cdot \bigl( (r_0 + h_0)^2 - r_0^2 \bigr)$\tcp*[r]{standard area}
  $r_\text{new} \gets r_0 + h_0$\;
  $h_\text{new} \gets \text{calculate\_normalised\_height}(r_\text{new}, A_\text{std}$)\;
  draw\_subtree($n, \; \theta_0, \; \beta_0, \; C_{ar0}, \; r_\text{new}, \; h_\text{new}, \; A_\text{std}, \; C_\text{acr}$)\;
\end{algorithm}

Algorithm \ref{alg:Main} outlines the top level of the RIT algorithm, where $n$ is the handler of the root node. $\theta_0$, $\beta_0$, $r_0$, and $h_0$ specify the annular sector for the root node. When $\theta_0 = 0, \; \beta_0 = 2\pi$, it is a full annulus. When $\theta_0 = 0, \; \beta_0 = 2\pi, \; r_0 = 0$, it is a circle with radius $h_0$. $C_{ar0}$ and $C_\text{acr}$ are two constants for controlling the double-wedge angle of each annular sector except for the root node. $C_{ar0}$ is the initial angle ratio $\alpha / \beta$, which is used for removing wedges in \texttt{draw\_subtree()}. $C_{ar0} > 0$ is the angle change rate ($>0$) that defines how the angle ratio $ar$ may change according to the tree depth. When $C_\text{acr}=1$, the angle ratio does not change, i.e., the initial value $C_{ar0} > 0$ will be used for all nodes except the root node. When $C_\text{acr}=0.9$, the angle ratio decreases gradually from the first generation of child-nodes towards the leaf-nodes, i.e., $C_{ar0}$ for children, $C_{ar0}C_\text{acr}$ for grandchildren, $C_{ar0}C_\text{acr}^2$ for great-grandchildren, and so on.
$C_{ar0}$ and $C_\text{acr}$ are typically defined as global constants.

Algorithm \ref{alg:Subtree} outlines a recursive procedure for processing a subtree. The procedure is invoked after the parent node $n$ has already been drawn. The procedure \texttt{draw\_subtree()} receives the following inputs from the proceeding calling procedure: 
\begin{itemize}
    \vspace{-1mm}
    \item $n$ --- a handler (e.g., an index or a pointer) of the parent node.
    \vspace{-1.5mm}
    \item $\theta$, $\beta$ --- the starting angle and the arc angle of the fan-shaped sector for accommodating this subtree.
    \vspace{-1.5mm}
    \item $ar$ --- the angle ratio $\alpha / \beta$ for the two wedges to be removed.
    \vspace{-1.5mm}
    \item $r$ --- the outer radius of the parent node $n$, which is the inner radius of its child nodes' annular sectors.
    \vspace{-1.5mm}
    \item $h$ --- the tentative height (i.e., without any top-up sector) of these child nodes' annular sectors.
    \vspace{-1.5mm}
    \item $A_\text{std}$ --- standard area for a 100\% annular sector, which is usually set based on the area of the root node of the whole tree.
    \vspace{-1.5mm}
    \item $C_\text{acr}$ --- angle change rate ($>0$) that defines how the angle ratio $ar$ may change according to the tree depth.
\end{itemize}

The procedure \texttt{draw\_subtree()} makes use of several subroutines, \revise{which are detailed in Appendix~\ref{app:Subroutine}.}

\begin{algorithm}
\caption{Recursive procedure: \texttt{draw\_subtree()}}\label{alg:Subtree}
  \KwIn{$n, \; \theta, \; \beta, \; ar, \; r, \; h, \; A_\text{std}, \; C_\text{acr}$}
  \tcp{DRAW all children's annular sectors}
  $\theta_c \gets \theta$\;
  \For(\tcp*[h]{for each child node}){$c_i \in n.\text{Children}$}{
    $c_i.\beta \gets 2 \cdot \pi \cdot c_i.\text{Data}$\;
    $c_i.\theta \gets \theta_c$\;
    draw\_a\_sector($c_i.\text{Color}, \; c_i.\theta, \; c_i.\beta \; r, \; h$)\;
    $\theta_c \gets \theta_c + c_i.\beta$\;
  }
  \tcp{REMOVE 2 wedges ($0.5\alpha$) for each child}
  \For(\tcp*[h]{for each child node}){$c_i \in n.\text{Children}$}{
    $c_i.\alpha \gets \text{set\_wedge\_angle}(ar, c_i.\beta)$\;
    remove\_wedge\_start($c_i.\theta, \; c_i.\alpha, \; r, \; h$)\;
    remove\_wedge\_end($c_i.\theta + c_i.\beta, \; c_i.\alpha, \; r, \; h$)\;
  }
  \tcp{ADD top-up sectors \& RECURSION for subtrees}
  \For(\tcp*[h]{for each child node}){$c_i \in n.\text{Children}$}{
    $\theta_t \gets c_i.\theta + 0.5 \cdot c_i.\alpha$\;
    $\beta_t \gets c_i.\beta - c_i.\alpha$\;
    $A_\text{wedge} \gets \text{calculate\_wedge\_area}(r, \; h, \; c_i.\alpha)$\;
    $r_t \gets r + h$\;
    $h_t \gets \text{calculate\_topup\_height}(r_t, \; c_i.\alpha, \; c_i.\beta, \; A_\text{wedge})$\;
    draw\_a\_sector($c_i.\text{Color}, \; \theta_t, \; \beta_t, \; c_i.\alpha, \; h_t$)\;
    $r_\text{new} \gets r_t + h_t$\;
    $h_\text{new} \gets \text{calculate\_normalised\_height}(r_\text{new}, A_\text{std}$)\;
    draw\_subtree($c_i, \; \theta_t, \; \beta_t, \; ar \cdot C_\text{acr}, \; r_\text{new}, \; h_\text{new}, \; A_\text{std}, \; C_\text{acr}$)\;
  }
\end{algorithm}

%% file: 6_testing.tex
\section{Testing and Results}
\label{sec:Testing}
In this section, we report the testing of the visual design and the algorithm of RIT in conjunction with synthetic data featuring the three issues associated with icicle and sunburst trees, open data in the public domain, and application-specific data. 

\begin{figure*}[th]
    \centering
    \begin{tabular}{@{}c@{\hspace{3mm}}c@{\hspace{3mm}}c@{\hspace{4mm}}c@{}}
         \raisebox{9mm}{\includegraphics[height=20mm]{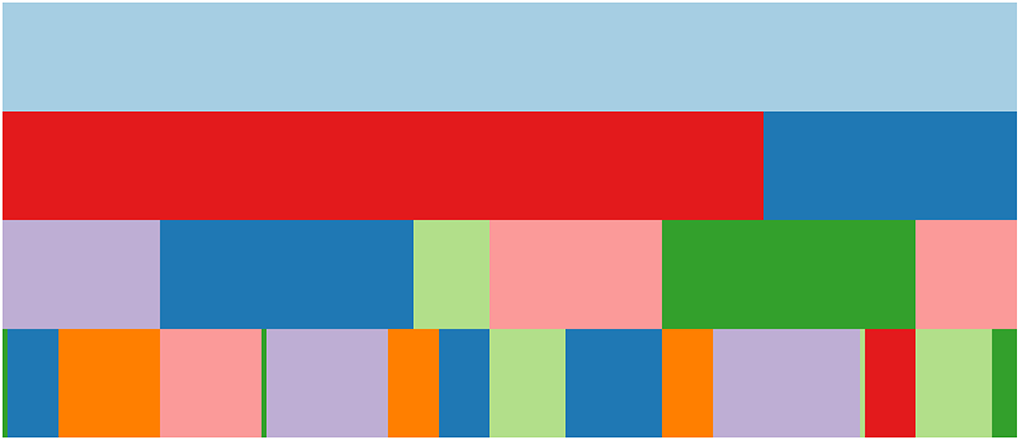}} &
         \includegraphics[height=38mm]{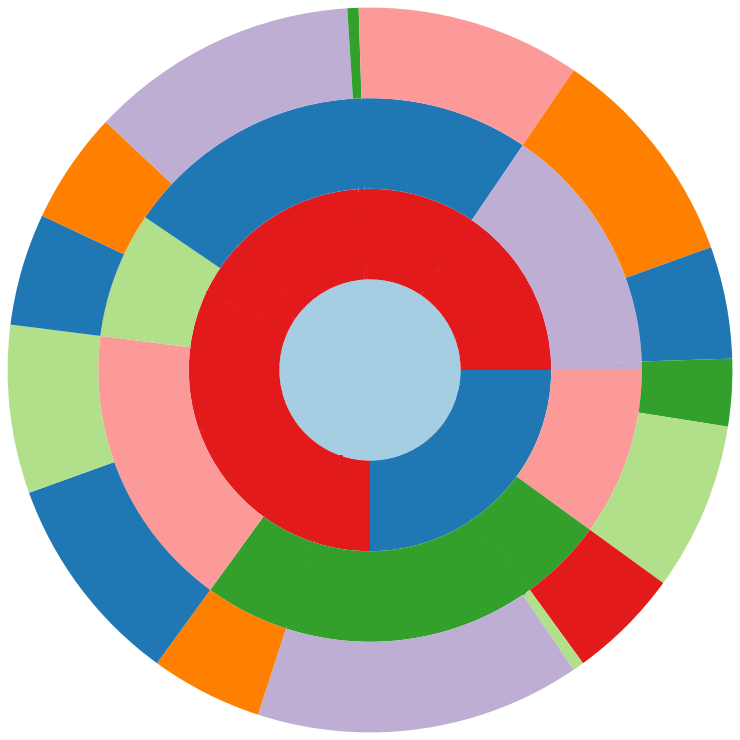} &
         \includegraphics[height=38mm]{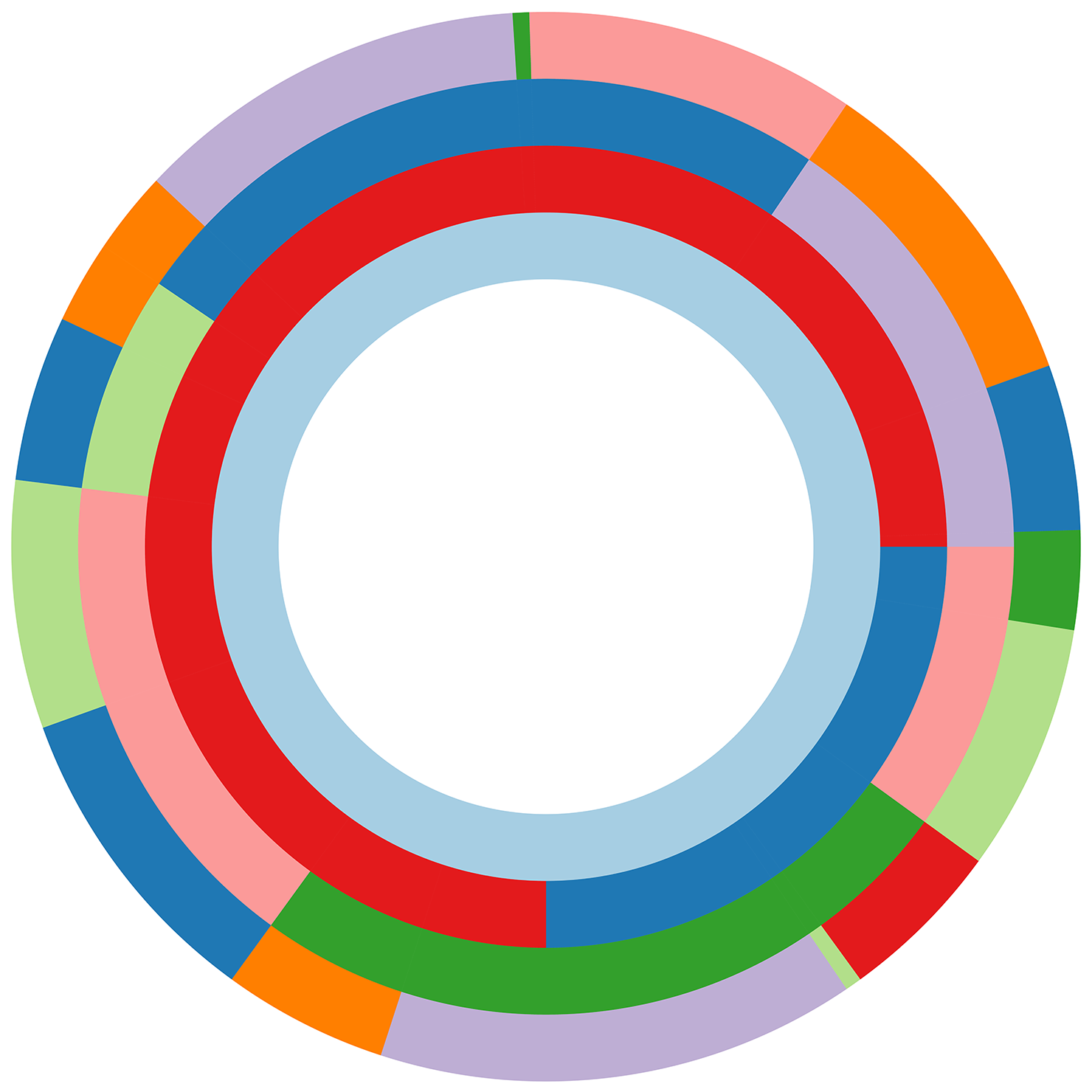} &
         \includegraphics[height=38mm]{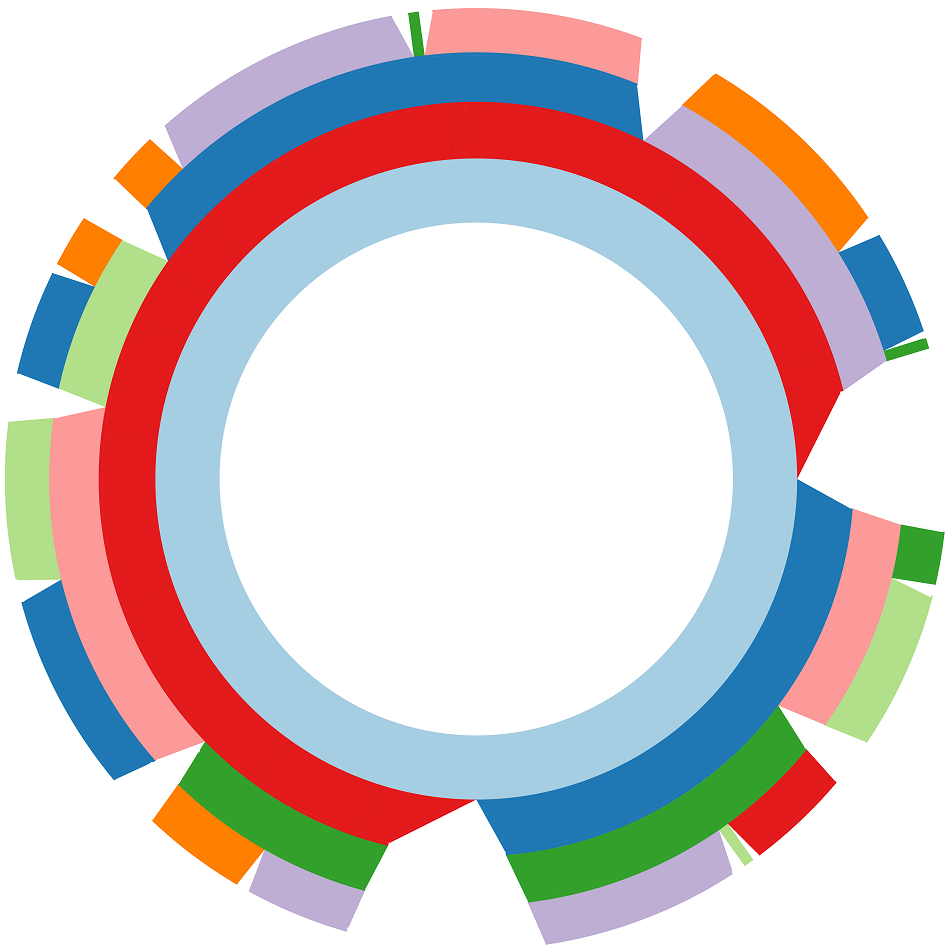} \\
         \small (a) Icicle tree &
         \small (b) Sunburst ($r_0=0, h=2$) &
         \small (c) Sunburst ($r_0=8, h=2$) &
         \small (d) RIT ($r_0=8, h_0=2$)\\[2mm] 
    \end{tabular}
    \begin{tabular}{@{}c@{\hspace{4mm}}c@{\hspace{2mm}}c@{}}
         \raisebox{3mm}{\includegraphics[height=30mm]{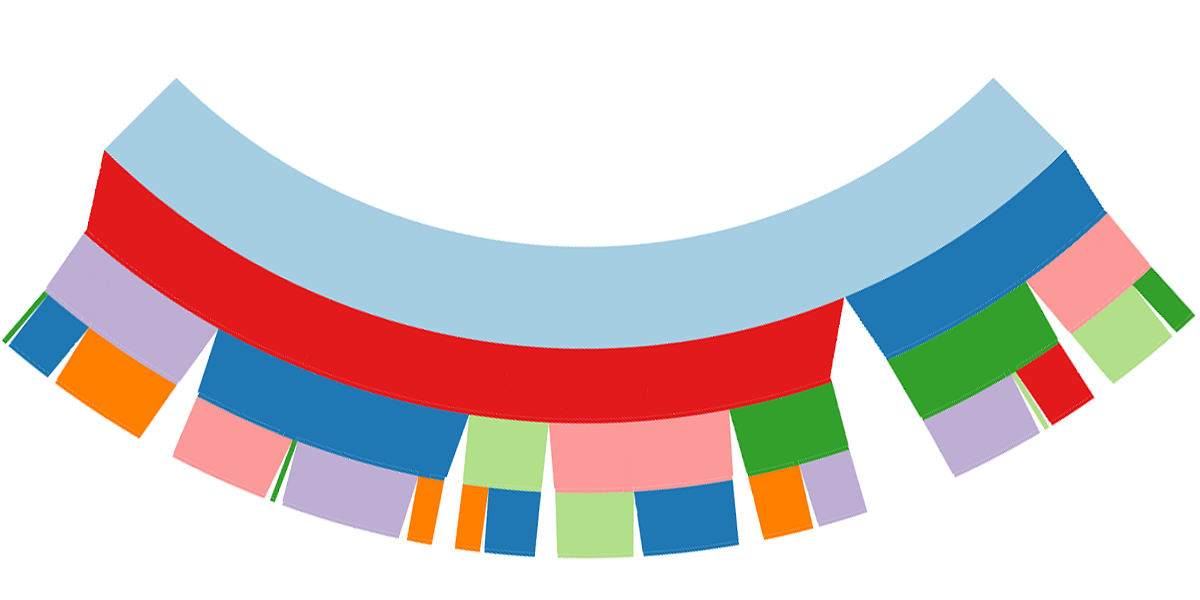}} &
         \raisebox{3mm}{\includegraphics[height=30mm]{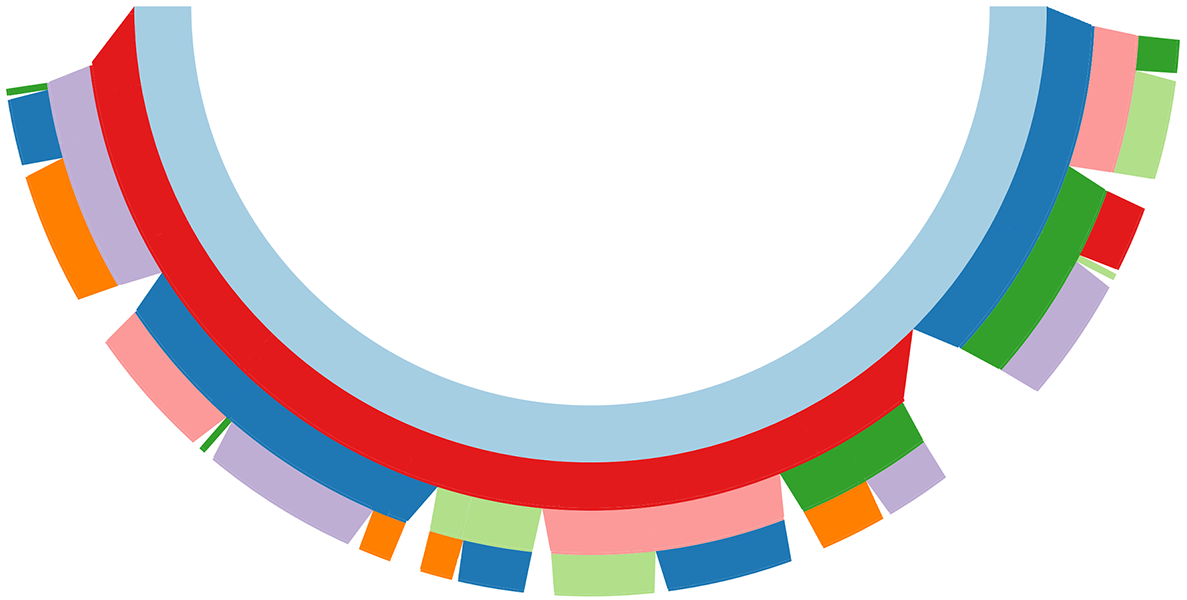}} &
         \includegraphics[height=36mm]{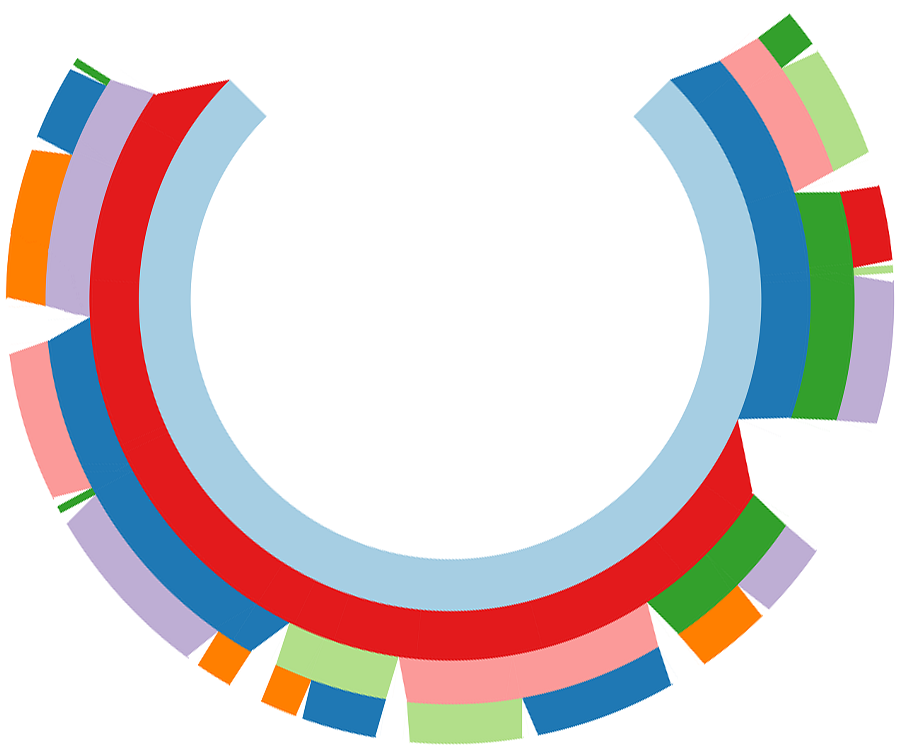} \\
         \small (e) RIT ($\theta_0=1.25\pi, \beta_0=0.5\pi, r_0=20.5, h_0=2$) &
         \small (f) RIT ($\theta_0=\pi, \beta_0=\pi, r_0=17.5, h_0=2$) &
         \small (g) RIT ($\theta_0=0.75\pi, \beta_0=1.5\pi, r_0=14.3, h_0=2$)\\[2mm]
    \end{tabular}
    \begin{tabular}{@{}c@{\hspace{4mm}}c@{\hspace{4mm}}c@{\hspace{4mm}}c@{}}
         \includegraphics[width=40mm]{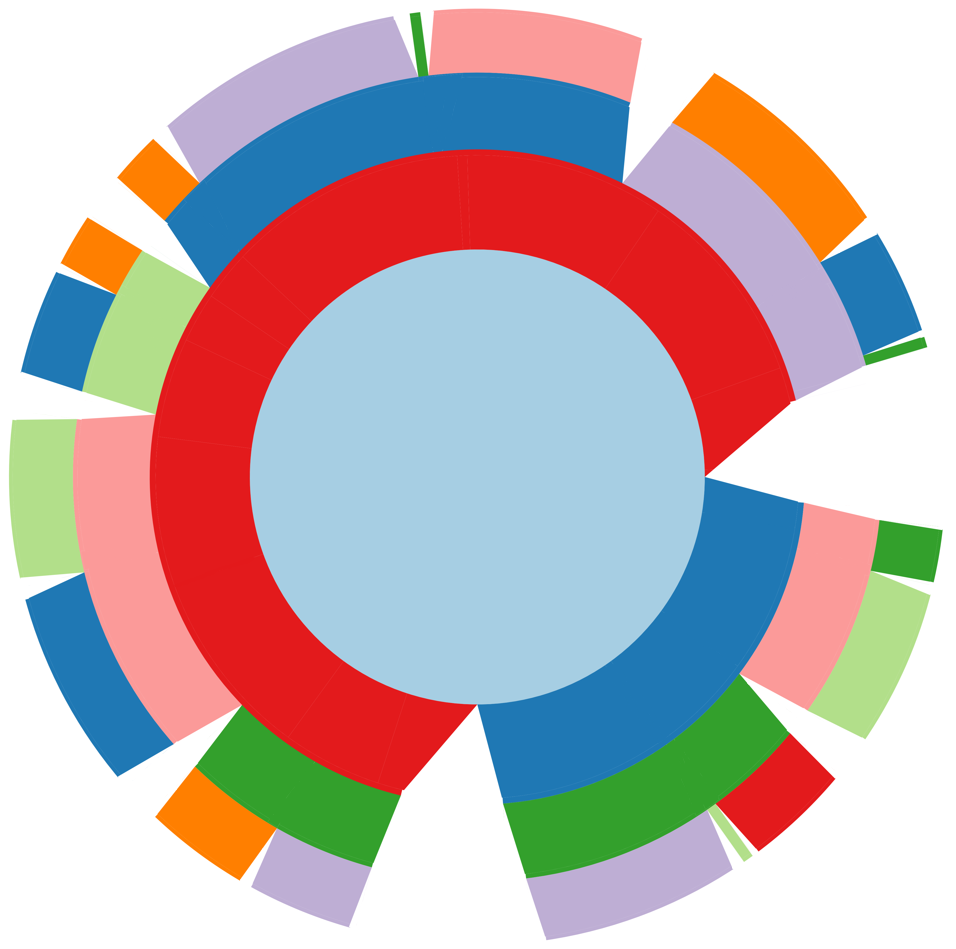} &
         \includegraphics[width=40mm]{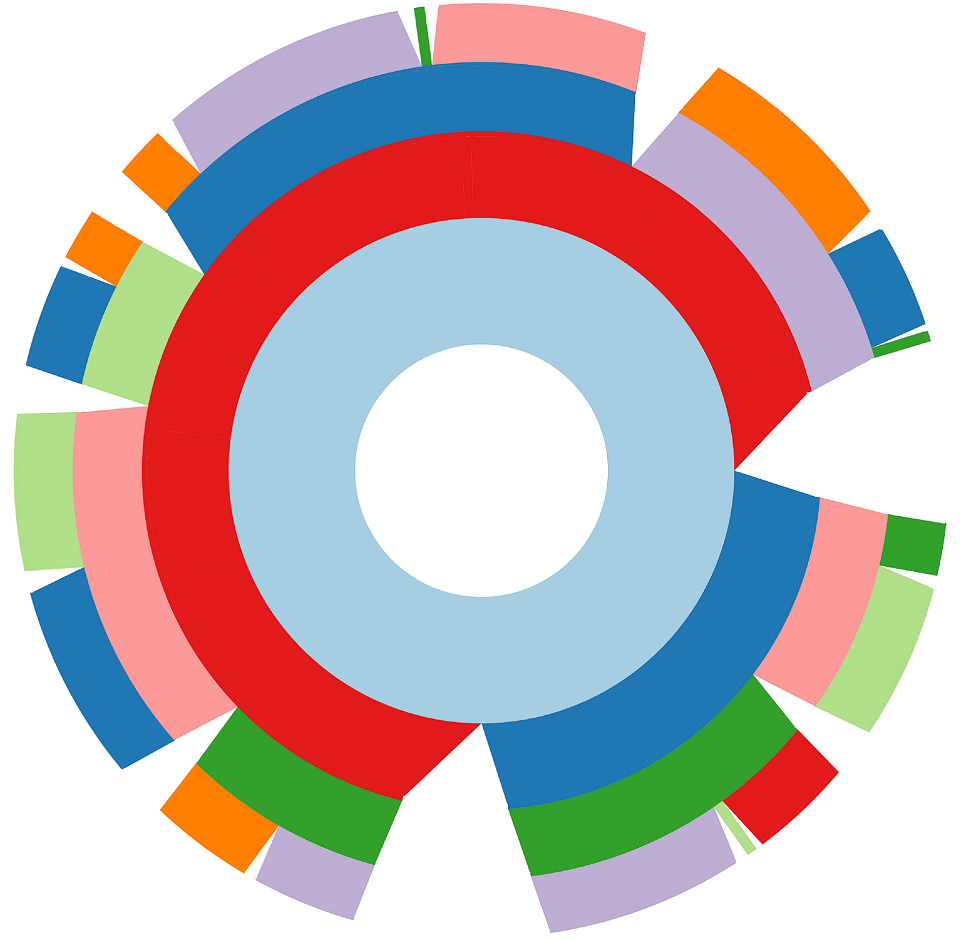} &
         \includegraphics[width=40mm]{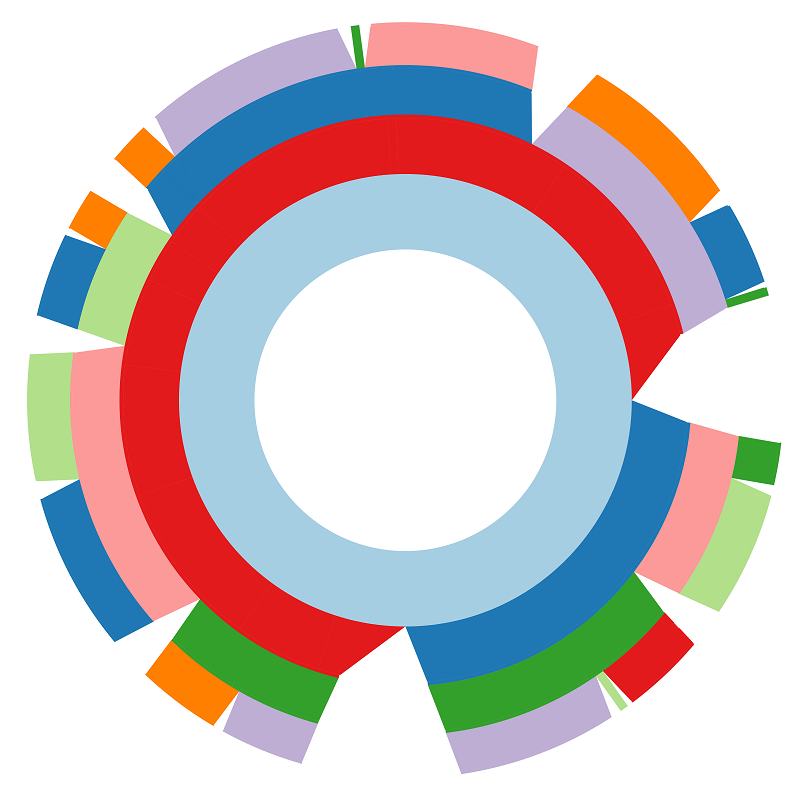} &
         \includegraphics[width=40mm]{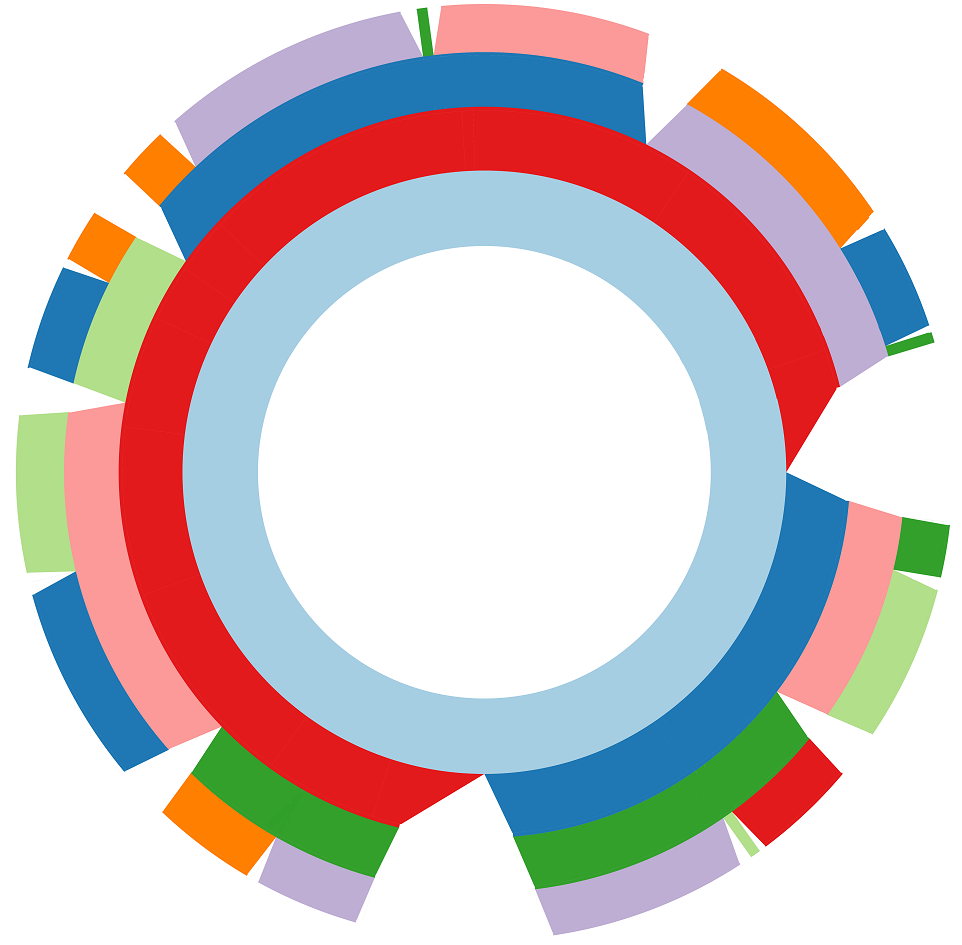} \\
         \small (h) RIT ($r_0=0, h_0=2$) &
         \small (i) RIT ($r_0=2, h_0=2$) &
         \small (j) RIT ($r_0=4, h_0=2$) &
         \small (k) RIT ($r_0=6, h_0=2$)
    \end{tabular}
    \caption{The dataset used to illustrate the three issues in Fig.~\ref{fig:Problems} was used to test RIT plots with different configuration parameters.}
    \label{fig:Synthetic}
    \vspace{-4mm}
\end{figure*}
\subsection{Problem-driven Testing}
We constructed several testing datasets that exhibit issues (a), (b), and (c) discussed in Section~\ref{sec:Intro}. This allowed us to control the complexity of a tree structure, the number of problematic nodes, the severity of the problems, where in the tree the problems occur, and so on. Fig.~\ref{fig:Problems} shows one such synthetic dataset. As annotated in Fig.~\ref{fig:Problems}, we assume that the color of each node encodes the categorical label of the node, and nodes with the same color are of the same category. The number in each node indicates its normalized data value such that the root is of the full amount of 100. The area of each node encodes the data value of the node. Fig.~\ref{fig:Synthetic} displays this synthetic dataset in different visual representations, i.e., (a) icicle tree, (b,c) sunburst trees, and (d$\sim$k) RITs. 

As we have already made observations about the three issues by comparing the three plots in Figs. \ref{fig:Synthetic}, here we focus on other observations. Firstly Fig.~\ref{fig:Synthetic}(e) can be seen as a slightly-curved version of an icicle tree. Issue (b) -- merged nodes -- has been resolved completely, while it does not suffer from the issue (c) -- inconsistent area mapping. Issue (a) -- thin nodes -- is significantly improved in comparison with Fig.~\ref{fig:Synthetic}(a). The three thin leaf nodes (two in green and one in pale green) are visible but not quite noticeable. Their distinguishability improves when the angle of the root sector $\beta_0$ (in radian) increases from $0.5\pi$ in (e) to $\pi$ in (f), then to $1.5\pi$ in (g), and $2\pi$ in (d). Comparing Figs. \ref{fig:Synthetic}(c,d), we can observe that the three thin nodes are more noticeable in (d) RIT than in (c) sunburst. Note that because of the need to ensure area constancy, we cannot enlarge thin nodes without enlarging other nodes. So nodes with small data values will always be relatively small unless one does not encode the data values using shape areas.

Similar to sunburst trees, RITs with a circular layout can adjust the inner radius of the root node, $r_0$, to change the size of the empty space at the center. Figs. \ref{fig:Synthetic}(h$\sim$k,d) illustrate the change of $r_0 = 0, 2, 4, 6, 8$. With sunburst trees, $r_0$ is commonly set to a large value, so the issue (c) -- inconsistent area mapping -- is less obvious. Since RITs do not suffer from the issue (c), the change of $r_0$ may serve different purposes. For example, a user may wish to leave plenty of blank space in the center for some text or a legend. Alternatively, a user may wish for each annular sector to have a relatively large $h_i$ for placing a text label.     

\begin{figure*}[th]
    \centering
    \begin{tabular}{@{}c@{\hspace{8mm}}c@{}}
         \includegraphics[height=86mm]{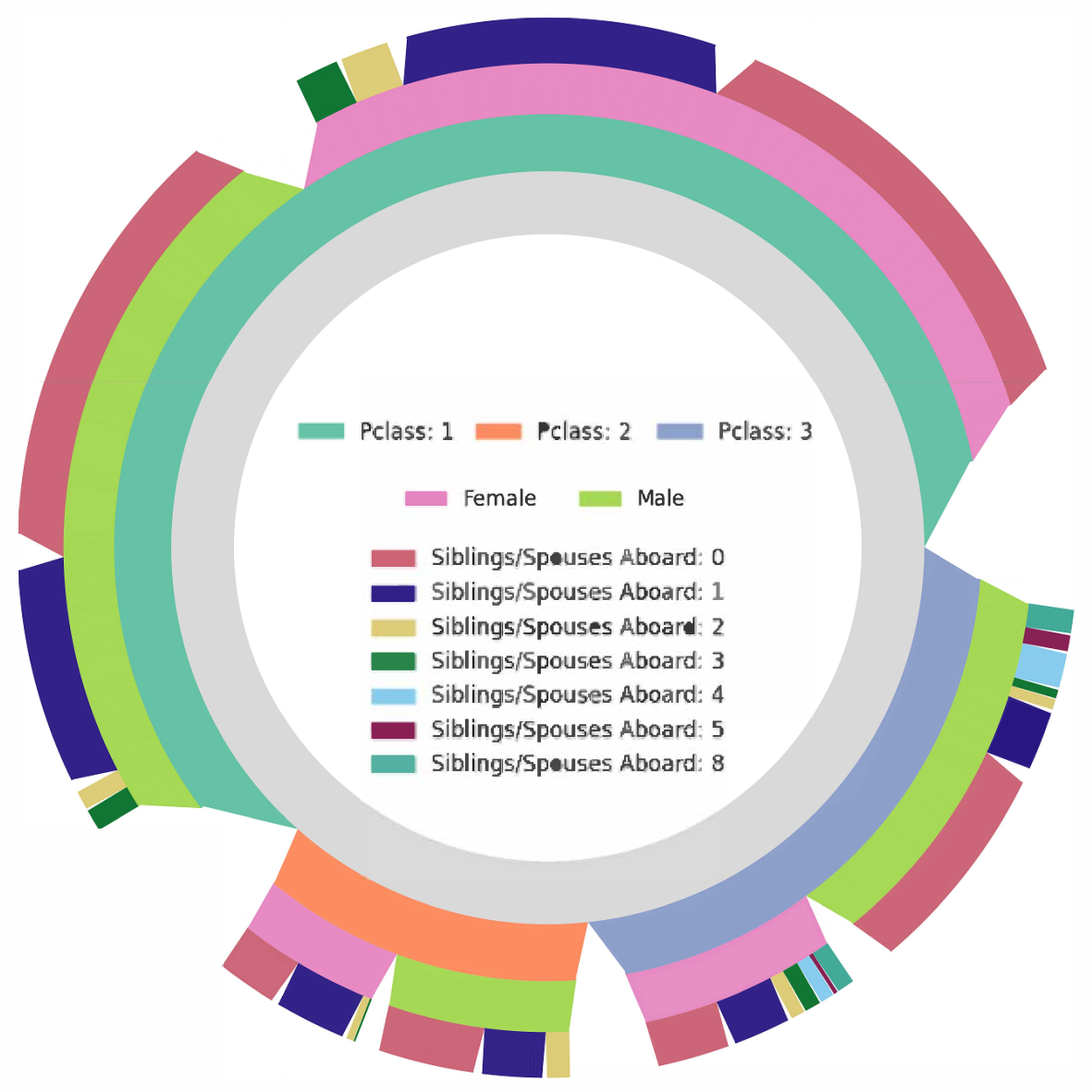} &
         \includegraphics[height=86mm]{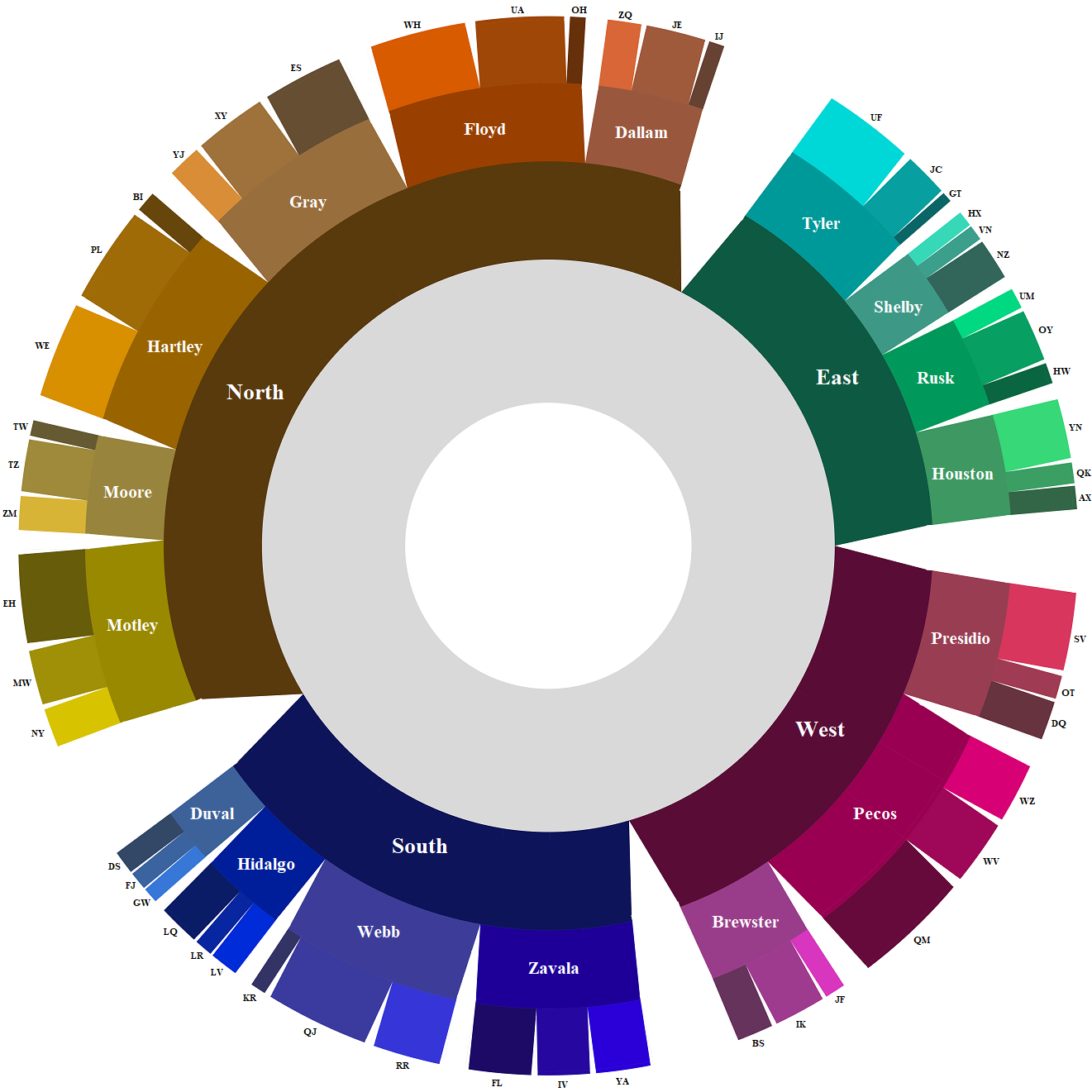} \\
         \small (a) Titanic passenger dataset (data source: \url{https://web.stanford.edu/class/}) &
         \small (b) The salesperson dataset (data source: \url{https://raw.githubusercontent.com/} \\
         \url{archive/cs/cs109/cs109.1166/stuff/titanic.csv} &
         \url{plotly/datasets/master/sales_success.csv})
    \end{tabular}
    \caption{RIT plots for two open datasets in the public domain. The Titanic dataset features a good number of thin nodes, while the salesperson dataset has many nodes that need to be identified using text labels as a color legend with many colors would not be appropriate.}
    \label{fig:Public}
    \vspace{-4mm}
\end{figure*}
\subsection{Testing with Datasets in the Public Domain}

We have also tested several open datasets in the public domain. Fig.~\ref{fig:Public} shows two example RITs produced using two of these datasets respectively. The Titanic passenger dataset contains some information about each passenger aboard the Titanic, including ticket class, gender, and a categorical label about how many accompanied persons. In Fig.~\ref{fig:Public}(a), the grey root annulus is the total number of passengers. At tree depth 2, passengers are clustered into three categories based on their ticket classes. At tree depth 3, passengers are then divided into male and female groups. At the leaf level, passengers are further divided into seven categories according to the number of accompanied persons. There are thin nodes in the tree, indicating that some passengers with ticket classes 2 and 3 were traveling with many accompanied persons. The RIT plot is able to handle these thin nodes. The data indicate no passenger who was traveling with 6 or 7 accompanied persons. Passengers with first-class tickets have 3 or fewer accompanying persons.

Fig.~\ref{fig:Public}(b) shows a dataset with four regions, each region of which has several counties.
Each leaf node is associated with an individual salesperson and the size of the node encodes the amount of sales values by the salesperson. We designed a colormap that allocates four distinct hue ranges to the four regions respectively. The colors of the counties and salespersons are algorithmic assigned such that the colors are of different HSV values, except that the hue value H is within a smaller hue-range defined by the color of the region. 

The gaps between nodes are more important in this RIT as the node colors can be rather similar. As we can observe from Fig.~\ref{fig:Public}(b), all nodes are well separated.

\begin{figure*}[th]
    \centering
    \begin{tabular}{@{}c@{\hspace{4mm}}c@{\hspace{4mm}}c@{\hspace{4mm}}c@{}}
         \includegraphics[width=40mm]{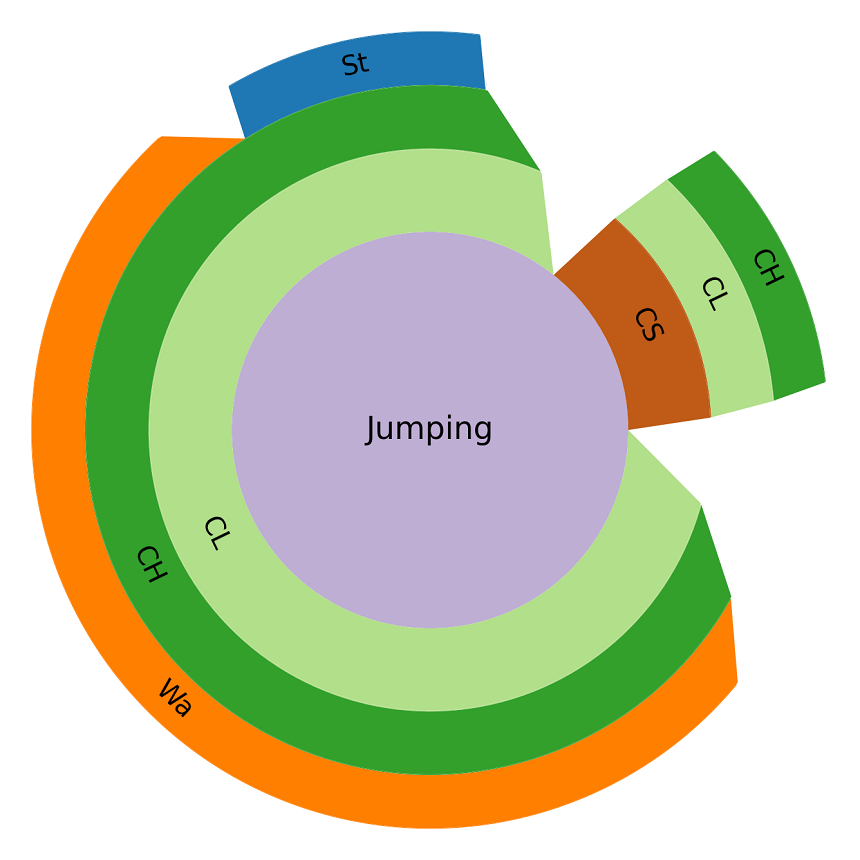} &
         \includegraphics[width=40mm]{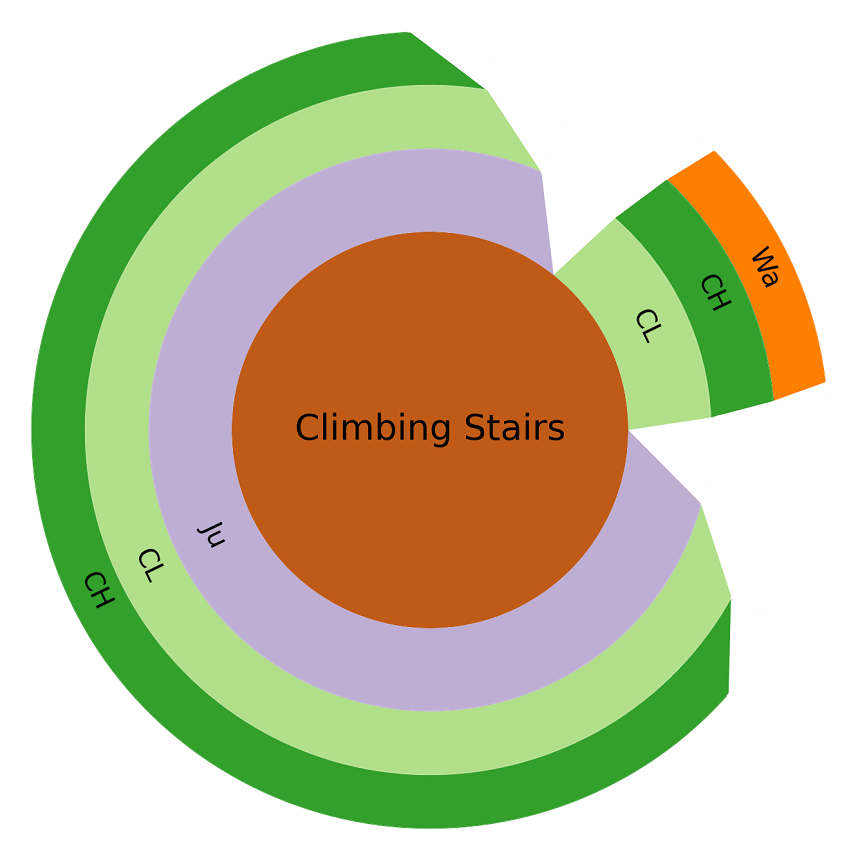} &
         \includegraphics[width=40mm]{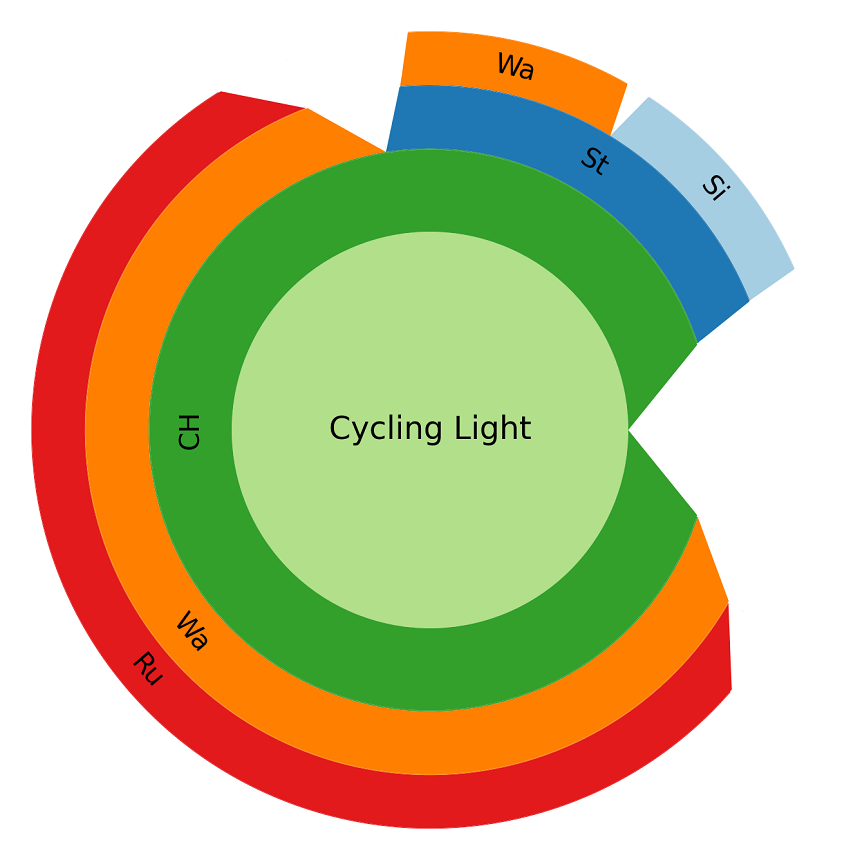} &
         \includegraphics[width=40mm]{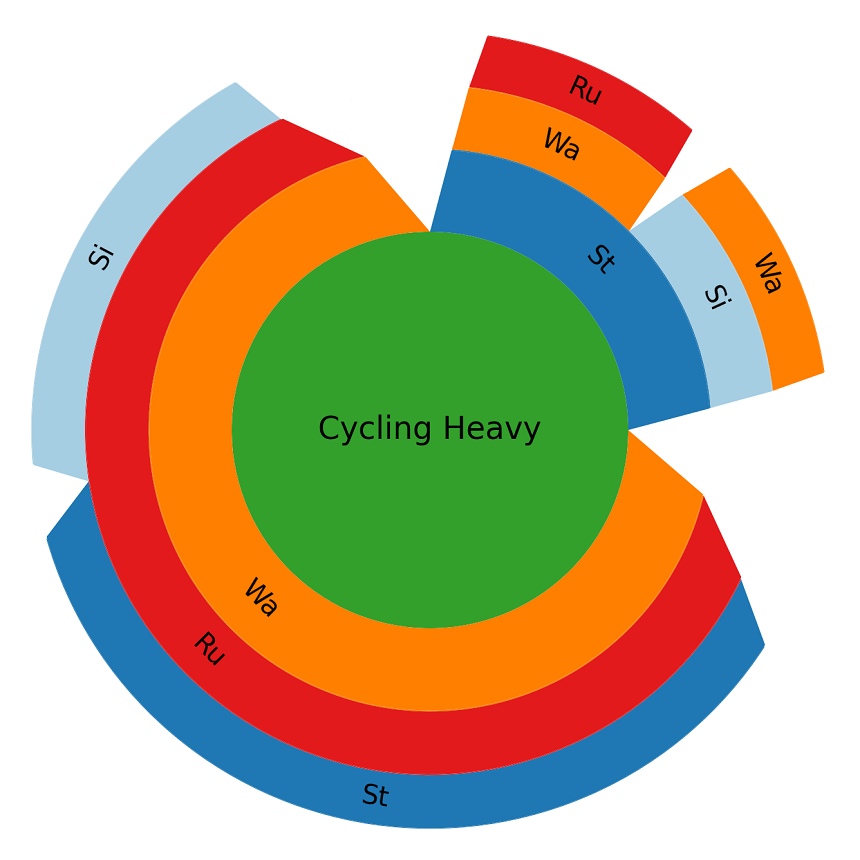} \\
         \includegraphics[width=40mm]{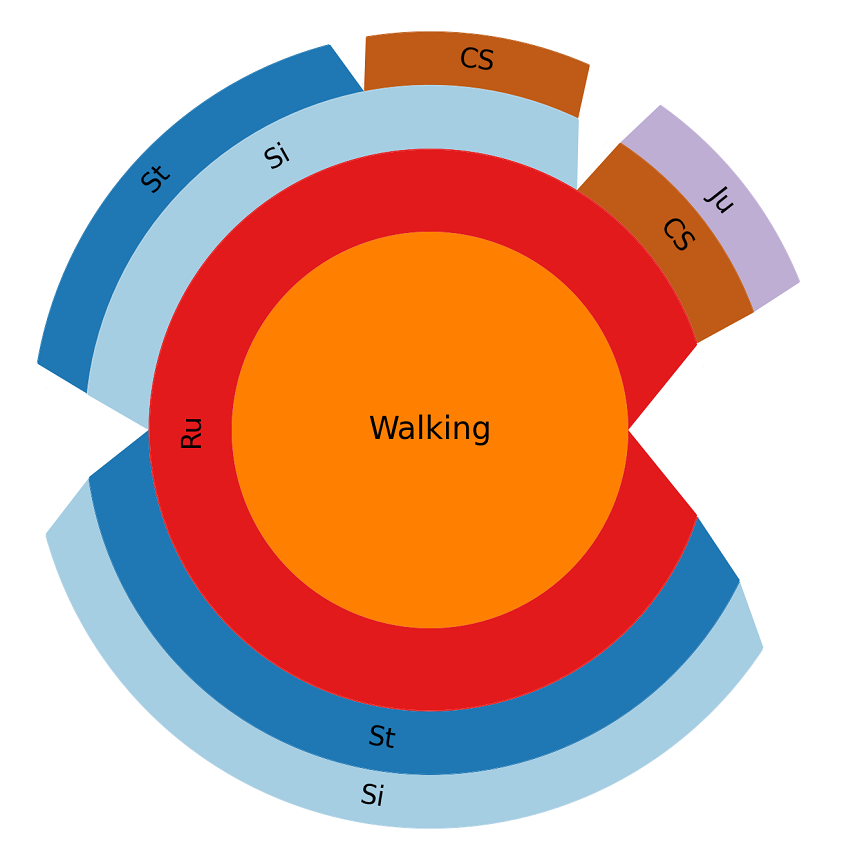} &
         \includegraphics[width=40mm]{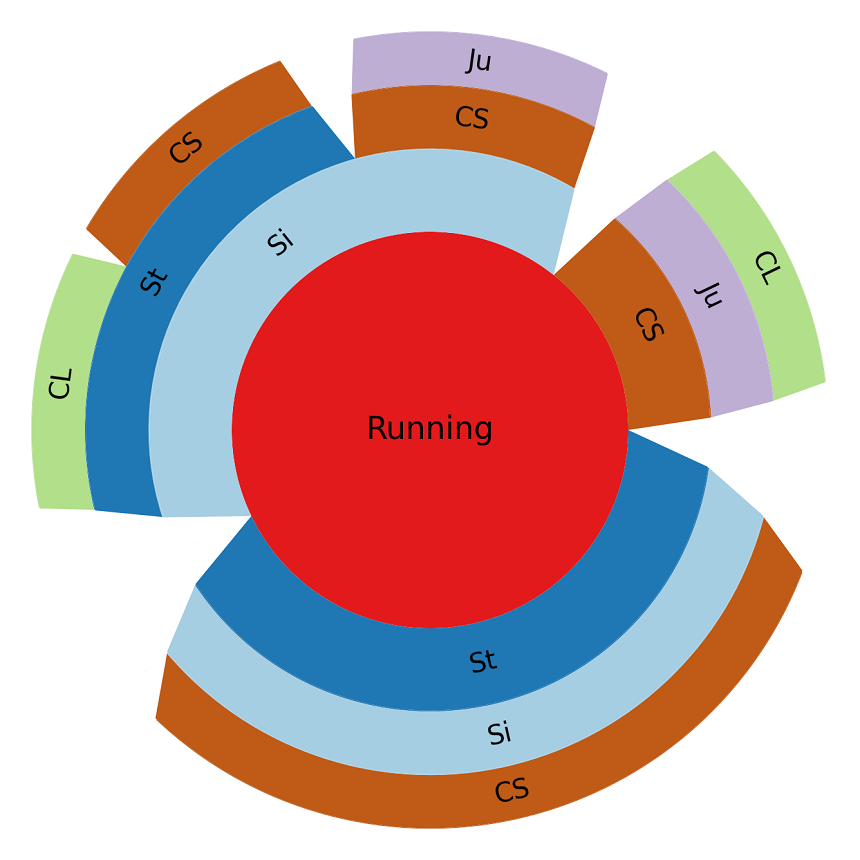} &
         \includegraphics[width=40mm]{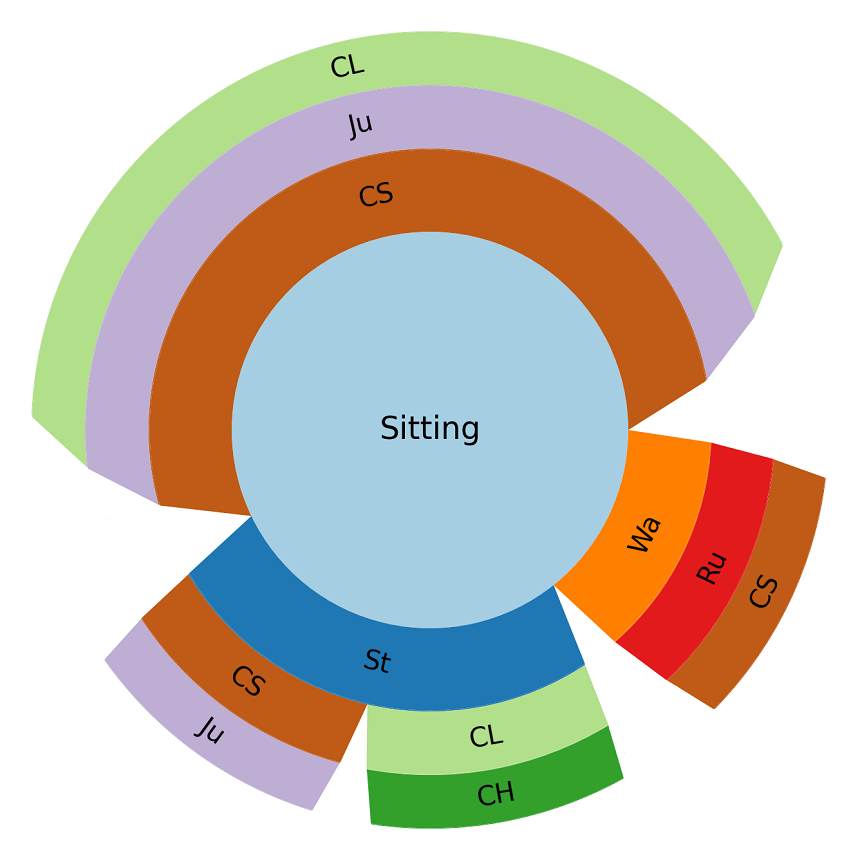} &
         \includegraphics[width=40mm]{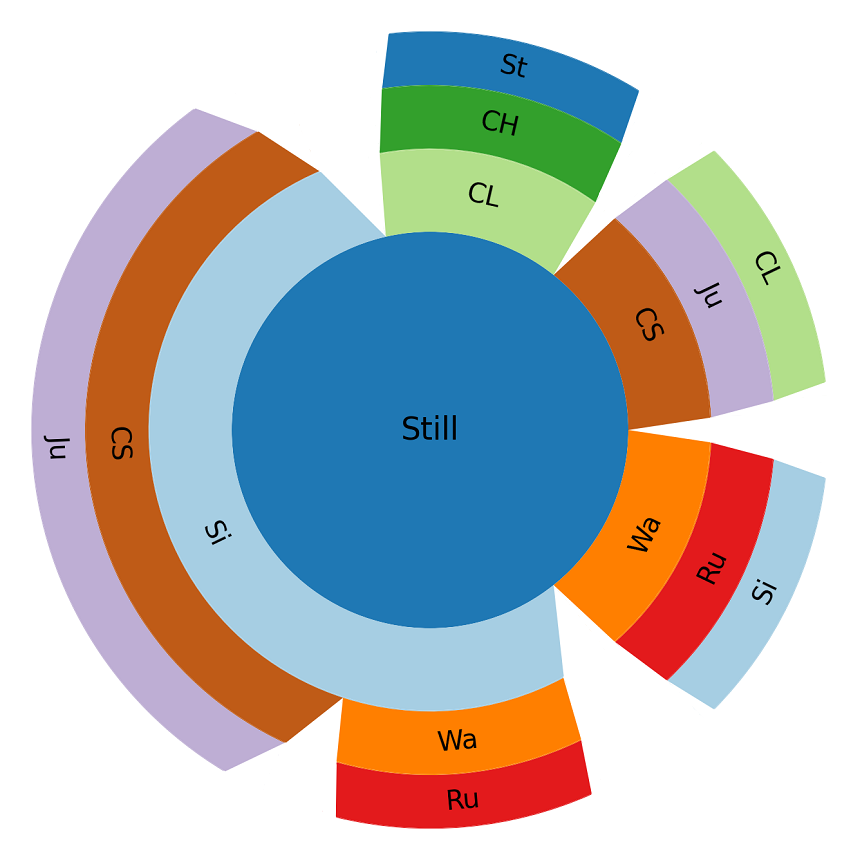} \\
    \end{tabular}
    \includegraphics[width=160mm]{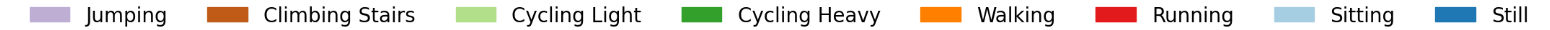}
    \caption{The statistics about three-step state transitions from each of the eight states to different states.}
    \label{fig:CBS}
    \vspace{-2mm}
\end{figure*}
\subsection{Testing with Data in a Specific Application}
\label{sec:CBS-data}
\revise{Statistics Netherlands collects, processes, and analyzes a variety of data that features tree or graph representations. There are two major categories of such data: (i) data represented as explicit tree structures (e.g., data pertaining to the geographical regions of a country, categories of land uses, performance indicators of economic sectors, and classifications of goods), and (ii) tree structures extracted by processing other types of data. For example, some data sets are collected as temporal data with categorical and numerical values, such as employment data in socioeconomics and movement data in public transport. In the context of the latter, Statistic Netherlands is particularly interested in state transitions, e.g., unemployed to employed or changes in transport modality. To enable statistical analysis of the relations among different events in the data, the original temporal data is commonly processed to extract relational information into tree or graph representations.}


To provide data about the general health and activity of the Dutch population, Statistics Netherlands regularly sends out surveys investigating the amount of physical activity in a representative sample of the Dutch population.
A problem with these surveys is that the measured activity levels can be subjective.
In general, self-reported activity levels are overestimated, which introduces subjectivity in health statistics.
In search of more objective measures, Statistic Netherlands, in cooperation with other public parties like the Dutch Health Authorities (RIVM), is investigating the use of accelerometers. Accelerometers measure the acceleration of a subject in three axes ($x$, $y$, $z$) and can be used to measure the intensity of
physical activity and moreover be used to give an estimation of the type of activity performed.
As such, accelerometer data can provide a more detailed view of physical activity during day-to-day life in the Dutch population.
Using accelerometers, Statistics Netherlands hopes to obtain more detailed statistics about the amount of moderate to vigorous physical activity (MVPA), but also insights into sedentary behavior and sleep patterns in the general Dutch population.

Statistics Netherlands has been studying the statistics of different events in the data collected from wearable devices. 
One particular analytical need is to observe, compare, and reason the statistics about the transitions among different events. For example, we may consider eight types of movement events, including \emph{climbing stairs}, \emph{cycling heavy}, \emph{cycling light}, \emph{jumping}, \emph{running}, \emph{sitting}, \emph{still}, and \emph{walking}. We can use RIT plots, such as those in Fig.~\ref{fig:CBS} to observe the statistics about transitions from event $x_1$ to event $x_2$, then event $x_3$, and so on.

 We first considered a number of visual representations for graphs, such as chord diagrams and Sankey diagrams. We found that it was difficult to compare different transition sequences in the same graph plot, e.g., to compare transitions:
 
\begin{tabular}{@{\hspace{8mm}}r@{\hspace{4mm}}l@{}}
         &  running $\Rightarrow$ sitting $\Rightarrow$ climbing stairs\\
     vs. & running $\Rightarrow$ climbing stairs $\Rightarrow$ sitting \\
     vs. & walking $\Rightarrow$ sitting $\Rightarrow$ climbing stairs \\
     vs. & walking $\Rightarrow$ climbing stairs $\Rightarrow$ sitting \\
\end{tabular}

The main challenge seemed to be the costly cognitive effort required for multiple observation tasks (i.e., locate one transition sequence and remember it, and then locate and remember another) before a comparison task can be performed. This made us consider the use of multiple tree representations instead of a single graph representation.

When each type of event has its tree, depicting the statistics of different transitions starting from the specific type of event, makes the location of the first event much easier, while depicting the following events in a much less cluttered way. The individual tree plots also provide external memorization. This reasoning led us to consider various tree visualization techniques that could encode statistical data with node sizes, which include treemap, icicle tree, sunburst tree, and their variants. We found that one seemed to need more cognitive effort to perceive the root and internal nodes in a treemap, and it was easier to notice the root and internal nodes with an icicle or a sunburst tree. However, we also identified some issues as illustrated in Fig.~\ref{fig:Problems}. This became part of the motivation for proposing a new visual design that could address these issues.

Fig.~\ref{fig:CBS} shows an example set of statistics for 3-transition sequences. The raw data were collected in a laboratory setting, and the main observational and analytical tasks were to compare the statistics of different transitional sequences \revise{(e.g., between cycle heavy and cycle light in Fig.~\ref{fig:CBS})} to see if they meet the anticipated statistical properties of the data collection exercise, and to compare with data collected in real-life and other laboratory settings.

\revise{In one of the evaluation meetings, domain experts confirmed the merits of RIT plots in comparison with the icicle tree and sunburst tree. With the mathematical assurance of the RIT plots, one can now compare different visual patterns without having to worry about the inconsistent size-encoding or mingled nodes. This can reduce the need for one to bring up numerical data after observing an interesting visual pattern. One domain expert specialized in machine learning also identified the uses of such visual comparison in other scenarios (e.g., comparing data collected in different regions, seasons, or times of day), and} would like to see the design to be deployed in a software system for supporting the modeling of movement data using machine learning.%

\revise{The evaluation meetings also resulted in a few suggestions for improvement, such as color mapping. The plots shown in Fig.~\ref{fig:CBS} are improved versions of the original plots discussed in two evaluation meetings. One domain expert, who is specialized in visualization but was not involved in the mathematical reasoning and algorithmic design, asked a series of technical questions as part of the evaluation. Some questions served as ``rigorous conceptual tests'' of the RIT design and algorithm, while other questions led to more in-depth discussions on several design decisions, including two questions in Section~\ref{sec:Discussion}.}  

%% file: 7_discussion.tex
\section{Discussions}
\label{sec:Discussion}

In this section, we discuss several questions about a few minor design decisions, which we encountered during our design, implementation, testing, and evaluation processes. \revise{In particular, the last question was posed by reviewers of this work.} 

%
\vspace{2mm}
\noindent\textbf{Setting a Shared Double-Wedge Angle?}
In our algorithm, the wedge to be removed has an angle ${\scriptstyle \frac{1}{2}\alpha}$ that is proportional to the angle $\beta$ of the corresponding annular sector. When the two neighboring nodes are of very different sizes, i.e., with very different $\beta$ angles, the two removed wedges form a noticeably-asymmetric shape between the two sectors. This led to the consideration of a possible design option for removing two wedges of the same angle. 

It was quickly realized that this ``shared double-wedge angle'' would have to be set according to the smaller neighboring node for each pair of sibling nodes. For a thin node (i.e., a very small $\beta$), the maximum rule discussed in Section~\ref{sec:MaxAlpha} (i.e., Eq.\,\ref{eq:MaxAlpha}) would ensure a very small $\alpha$. On balance, we considered that it is more important to have sufficient gaps near thin nodes than symmetry, as the former affects perception, while the latter affects aesthetics. We, therefore, selected the option of defining the double-wedge angle $\alpha$ proportional to the annular sector angle $\beta$ in our implementation.   

%
\vspace{2mm}
\noindent\textbf{Optimizing the Height of an Annular Sector?}
We also considered other options for determining the wedge angle and the size of the top-up sector. In Section~\ref{sec:Math}, we outlined one option, with which the double wedge angle $\alpha$ is determined first, and the attributes of the top-up sector (i.e., sector angle ($\beta - \alpha$) and height $h_\text{topup}$) are calculated accordingly. Mathematically, it is possible to determine $h_\text{topup}$, and then calculate $\alpha$. However, since the algorithm would have to validate $\alpha$ according to the maximum rule, any adjustment to $\alpha$ would lead to a change of $h_\text{topup}$, causing a looped calculation step.

Another option is to have an adjustable height $h$ for each annular sector, i.e., to adjust $h$ (instead of adding a top-up) after removing the two wedges. Aesthetically, the advantage is to have a straight line for the removed area. One would need to determine $\alpha$ and $h$ simultaneously, using an optimization technique, according to a set of predefined constraints for the size of the two wedges and the overall node size (i.e., the uncut annular sector $-$ two wedges). As an optimization process may lead to noticeable computational cost and implementation complexity, we hope that some future work may explore this option further.

\begin{figure}[th]
    \centering
    \begin{tabular}{@{}c@{\hspace{2mm}}c@{\hspace{2mm}}c@{}}
         \includegraphics[width=28mm]{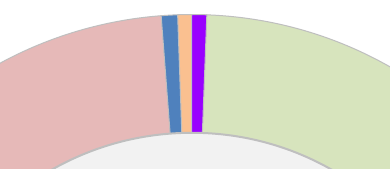} & 
         \includegraphics[width=28mm]{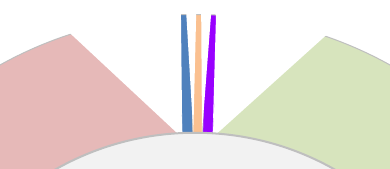} & 
         \includegraphics[width=28mm]{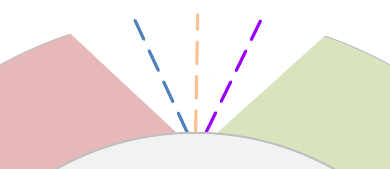} \\
         \small (a) three thin nodes &
         \small (b) after wedge removal &
         \small (c) space relaxation  
    \end{tabular}
    \caption{One practical solution for dealing with small gaps between thin nodes (i.e., with very small data values) is to relax the mathematically-defined gaps by redistributing the thin nodes in a large space that contains some large gaps. Meanwhile, it is necessary to modify the visual representations of these redistributed nodes to indicate the uncertainty of the encoded sizes.}
    \label{fig:Relaxation}
    \vspace{-4mm}
\end{figure}
%
\vspace{2mm}
\noindent\textbf{Dealing with Multiple Thin Nodes?}
As discussed previously, it is necessary to set $\alpha$ (double wedge angle) as a portion of $\beta$ (annular sector angle). However, the gaps between nodes can sometimes be rather small if two or more thin nodes are positioned next to each other as illustrated in Fig.~\ref{fig:Relaxation}. If an RIT plot is rendered into a scalable vector-graphics representation (e.g., SVG), the gaps will always be visible when one zooms into a part of an RIT plot containing thin nodes.

In many situations, an RIT plot may be rendered into a raster image, or users do not need to figure out the sizes of these thin nodes precisely though they do need to perceive the presence of these thin nodes. This may not strictly be a mathematical problem, but rather a common practical challenge that many different visual representations would face, including icicle trees and sunburst trees.

One practical solution is to relax the mathematically-defined gaps by redistributing each set of \emph{tightly-packed nodes} (TPNs) in a large space that contains some large gaps. This can be done using the following algorithmic steps:
\begin{enumerate}
    \item Identify each set of neighboring child nodes with small data values (i.e., less than a pre-defined threshold) as a TPN set, label these nodes in their records as nodes not to be displayed normally, and create a group identity for each TPN set such that the nodes in a TPN set can be processed together.
    \vspace{-1mm}
    \item For each TPN set, identify the total space available for redistributing the nodes in the TPN set. As illustrated in Fig.~\ref{fig:Relaxation}(b), typically a TPN set may have two neighboring nodes that are not TPNs and have left larger gaps. The total space thus includes these gaps as well as the space that would be occupied by the TPN set. In the case of Fig.~\ref{fig:Relaxation}(b), it is the space between the pale red node and the pale green node.\\
    $~$\hspace{5mm} When a TPN set includes the first or last child node of a parent node, at least one side of a TPN set does not border a large gap left by a sibling node. In such a case, we can use the wedge angle (${\scriptstyle \frac{1}{2}}\alpha$) of the parent node as we can infer that there is always a gap between the parent node and the parent's siblings.
    \vspace{-1mm}
    \item Redistribute the nodes in the TPN set evenly in the available space as illustrated in Fig.~\ref{fig:Relaxation}(c).
\end{enumerate}     

Meanwhile, it is necessary to modify the visual representations of these redistributed TPNs to indicate the uncertainty of the encoded sizes. We recommend using dashed lines, which are visually different from the ordinary nodes in an RIT, do not consume much space, and convey size uncertainty.

\begin{figure}
    \centering
    \includegraphics[width=87mm]{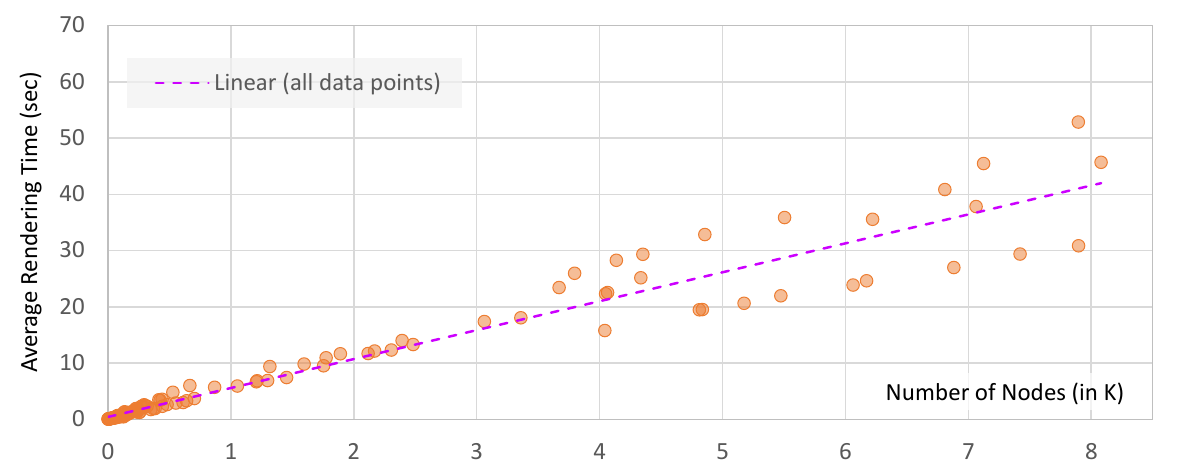}
    \caption{The results of a scalability test evidence that the RIT algorithm is linearly scalable in relation to the total number of nodes in a tree $N_v$. }
    \label{fig:Scalability}
    \vspace{-4mm}
\end{figure}

\vspace{2mm}
\noindent\revise{\textbf{How does the algorithm scale?} An analysis of the RIT drawing algorithm in Section \ref{sec:Algo} indicates that all nodes are visited three times (Algorithm \ref{alg:Subtree}). This suggests that the runtime complexity of the algorithm is O($N_v$) where $N_v$ is the total number of nodes. In other words, the algorithm is linearly scalable in relation to the number of nodes. Fig. \ref{fig:Scalability} shows the results of a scalability test that evidence the scalability analysis. More results of the test are given in Appendix \ref{app:Scalability}.}

%% file: 8_conclusion.tex
\section{Conclusions}
\label{sec:Conclusions}
In this paper, we have presented a new visual design, \emph{Radial Icicle Tree} (RIT), which addresses three issues identified in Section~\ref{sec:Intro} and illustrated in Fig.~\ref{fig:Problems}. Using the design process reported in Section~\ref{sec:Overview}, we were able to explore and analyze a number of design options systematically and identify the most promising design to focus on in the following mathematical and algorithmic considerations. Through our mathematical reasoning in Section~\ref{sec:Math}, we were able to confirm the desired mathematical properties of the design while working out a computational means for ensuring node separation and area constancy. We have presented a recursive algorithm for plotting RITs and validated the algorithm through an implementation in Python. We have tested the proposed visual design, its algorithm, and its implementation with synthetic data featuring all three issues concerned, open data in the public domain, and movement data that is being studied in Statistics Netherlands. The domain experts confirmed the merits of RIT in comparison with other visual designs that are currently being developed as part of a software system for supporting their analysis of movement data.  

Through the processes of visual design, mathematical reasoning, algorithm design, implementation, and testing, we have ascertained that RIT does address three issues without incurring any relative demerit. Through our discussion in Section~\ref{sec:Discussion}, we have identified a few limitations, which RIT has inherited from icicle and sunburst trees but offered potential opportunities for making further improvement through research and development. \revise{We have made our Python-based implementation available as open-source software at \url{https://github.com/MattJin19/Radial-Icicle-Tree}}. We also plan to develop an open-source implementation in D3.js.

%% file: 9_AppendixA.tex

\begin{center}
\large
APPENDICES OF\\[1mm]
\Large\noindent
\textbf{\textsf{Radial Icicle Tree (RIT):\\
Node Separation and Area Constancy}}\\[2mm]
\normalsize
Yuanzhe Jin$^1$, Tim J. A. de Jong$^2$, Martijn Tennekes$^2$, and Min Chen$^1$\\[1mm]
$^1 $University of Oxford and $^2 $Statistics Netherlands
\normalsize
\end{center}

\section{Further Details of the RIT Drawing Algorithm}
\label{app:Algorithm}

In this appendix, we provide further details of the RIT algorithm presented in Section \ref{sec:Algo}.

\subsection{The Main Data Fields in a Node Record}
\label{app:NodeRecord}
Each node record consists of the following data fields:

\begin{itemize}
    \vspace{-1mm}
    \item $n.Children$ --- a set of node handlers linking to all child nodes of node $n$. These child nodes may be stored in an array or a linked list. We denote them simply as $c_1, c_2, \ldots, c_i, \ldots \in n.Children$.%
    \vspace{-1mm}
    \item $n.Data$ --- a normalized data value to be mapped to the size of a node in a tree plot. Here we assume that the value is in the range of [0, 1], and $n.Data = 1$ for the root node indicating that it is mapped to 100\% of the standard area $A_\text{std}$. Typically, we may draw a root node as a full annulus defined by its inner radius $r_0$ and height $h_0$. We can use Eq.\,\ref{eq:AnnulusArea} to obtain its area $S_0$ and define $A_\text{std} = S_0$. Alternatively, we may draw the root node as an annular sector that starts at a polar angle $\theta$ and ends at an angle $\theta + \beta$. Here we specify angles in radian and confine $\beta$ to $0 < \beta \leq 2\pi$. $\theta$ is  normally specified in the range $[0, 2\pi]$, though it is not difficult for an algorithm to deal with any finite value of $\theta$ and converts it to an angle within the range $[0, 2\pi]$ whenever necessary.%
    \vspace{-1mm}
    \item $n.Color$ --- the color of node $n$. Here we assume that the color of each node has already been assigned for simplicity and clarity in describing the algorithm. In a practical implementation, the tree data structure will likely store application-specific data (e.g., type, quality level, etc.), which is mapped to color during the drawing process.
    \vspace{-1mm}
    \item $n.\theta$, \; $n.\beta$, \; $n.\alpha$ --- the starting angle, the arc angle, and the double-wedge angle of an annular sector corresponding to node $n$. These are computed dynamically by the algorithm, except that those for the root node are pre-defined before invoking \texttt{draw\_rit()}. In fact, it is not necessary to store these in the data record of $n$, except that it is slightly safer and more convenient to use these fields than internal variables in a recursive algorithm.  
\end{itemize}

\subsection{The Main Subroutines in the RIT Algorithm}
\label{app:Subroutine}

The main procedure \texttt{draw\_subtree()} in Algorithm \ref{alg:Subtree} (Section \ref{sec:Algo}) makes use of several subroutines, including:

\begin{itemize}
    \vspace{-1mm}
    \item \texttt{draw\_a\_sector()} --- This calls low-level one or more graphics subroutines to draw an annular sector.
    \vspace{-1.5mm}
    \item \texttt{set\_wedge\_angle()} --- This first determines $\alpha$ as a portion of $\beta$ according to the angle ratio $ar$, i.e., $\alpha \gets ar \cdot \beta$. It then checks to see if $\alpha$ violates the two constraints according to Section \ref{sec:MaxAlpha}. If $\alpha$ does, it adjusts $\alpha$ to meet the maximum rule.
    \vspace{-1.5mm}
    \item \texttt{remove\_wedge\_start()} --- This removes a wedge (${\scriptstyle \frac{1}{2}}\alpha$) at the starting end of an annular sector (see Section \ref{sec:NodeSeparation}).
    \vspace{-1.5mm}
    \item \texttt{remove\_wedge\_end()} --- This removes a wedge (${\scriptstyle \frac{1}{2}}\alpha$) at the terminating end of an annular sector (see Section \ref{sec:NodeSeparation}).
    \vspace{-1.5mm}
    \item \texttt{calculate\_wedge\_area()} --- This calculates the area of two wedges to be removed as described in Eq.\,\ref{eq:WedgeArea} in Section \ref{sec:NodeSeparation}.
    \vspace{-1.5mm}
    \item \texttt{calculate\_topup\_height()} --- This calculates the height of a top-up annular sector as described in Eq.\,\ref{eq:TopupHeight} in Section \ref{sec:NodeSeparation}. Note that the term $(r+h)$ is replaced with $r_t$ in \texttt{draw\_subtree()}.
    \vspace{-1.5mm}
    \item \texttt{calculate\_normalised\_height()} --- This calculates the height of an annular sector as described in Eq.\,\ref{eq:AnnulusHeight} in Section \ref{sec:AreaConstancy}. It does not take into account the area loss due to wedge removal, which will be compensated by the top-up annular sector.
\end{itemize}

%% file: 10_AppendixB.tex
\section{Scalability Test}
\label{app:Scalability}
We have conducted a scalability test to evaluate the conclusion of the theoretical analysis that the complexity of the RIT algorithm is linear, i.e., O($N_v$), where $N_v$ is the total number of nodes in a tree. We have created 112 trees of depths between 1 and 8. Each parent node may have up to 12 child nodes. We used three types of tree-generation algorithms:

\begin{itemize}
    \vspace{-1mm}
    \item \textbf{Fixed} ---
    Let $C_\text{max} > 1$ be the maximal number of child nodes per parent. This algorithm was used for relatively small $C_\text{max}$ values. Otherwise the total number of nodes grows too quickly in relation to the tree depth.
    \vspace{-1mm}
    \item \textbf{Random} ---
    For each parent node, the algorithm randomly selects a number $R_C \in [1, C_\text{max}]$, and creates $R_C$ child nodes. We noticed that for a slightly large value of $C_\text{max}$, the total number of nodes can also grow quickly in relation to the tree depth. This algorithm was used for trees that are not so deep or $C_\text{max}$ is not so large.%
    \vspace{-1mm}
    \item \textbf{Semi-random} --- We assign different values of $C_\text{max}$ at different tree depths with an overall decreasing trend from depth 2 to leaf nodes though $C_\text{max}$ may not necessarily decreases at each depth. For each parent node, the algorithm creates child nodes in the same way as the above \textbf{Random} algorithm.
\end{itemize}

\begin{figure}[t]
    \centering
    \begin{tabular}{@{}c@{\hspace{4mm}}c@{}}
    \includegraphics[width=39mm]{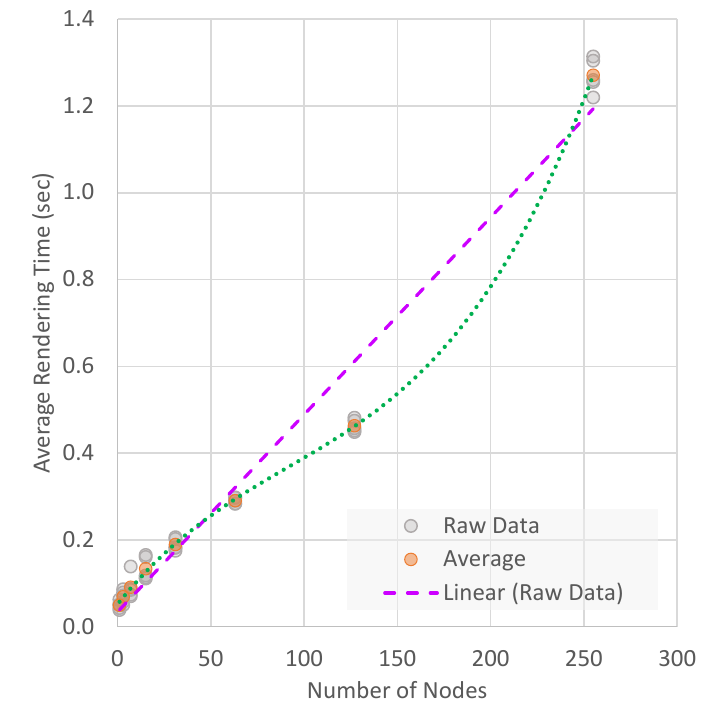} &
    \includegraphics[width=39mm]{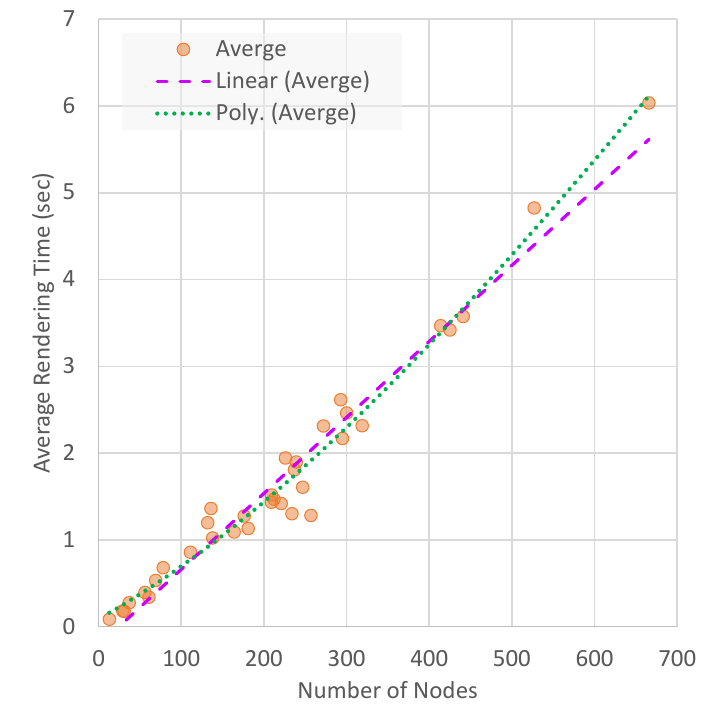} \\
    \small (a) Fixed: $C_\text{max}=2$ &
    \small (b) Semi-random: $C_\text{max}=12$\\[2mm]
    \includegraphics[width=39mm]{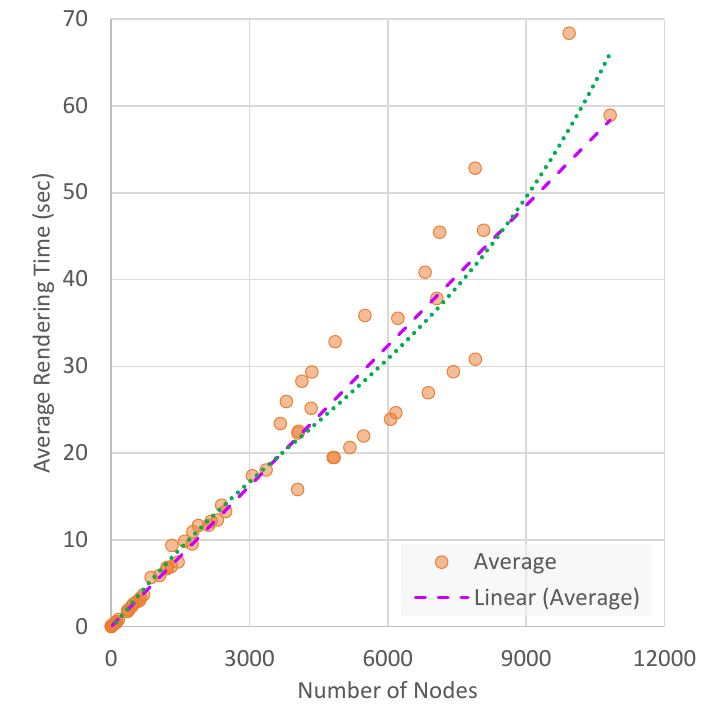} &
    \includegraphics[width=39mm]{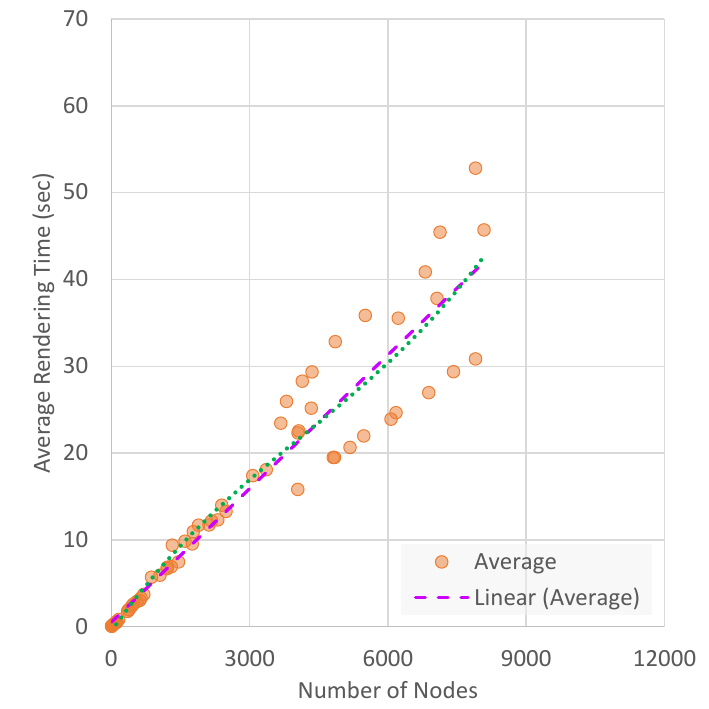}\\
    \small (c) Random: $C_\text{max}=8$ &
    \small (d) Remove two data points in (c)
    \end{tabular}
    \caption{Scalability tests for three different sets of trees.}
    \label{fig:S-Data}
    \vspace{-4mm}
\end{figure}

Applying the RIT algorithm to trees generated using the above three tree-generation algorithms exhibited similar scalability patterns but not exactly the same. Fig. \ref{fig:S-Data}(a) shows the scalability test of eight binary trees, which are of different depths and where each parent node has exactly two child nodes except the leaf nodes. Each tree was tested five times, and its average rendering time (sec.) is shown as a single orange dot. The cubic polynomial trend line (green dotted) fits the average data points almost perfectly and differs from the linear trend line (purple dashed) noticeably. For semi-randomly created trees with $C_\text{max}=12$, the difference between cubic and linear trend lines is reduced as shown in Fig. \ref{fig:S-Data}(b). For randomly-generated trees with $C_\text{max}=8$, there seems to be a similar difference between cubic and linear trend lines as shown in Fig. \ref{fig:S-Data}(c). However, a close look indicates that the sampling for trees with more than 9000 nodes was too sparse to be reliable. In Fig. \ref{fig:S-Data}(d), we removed the two trees with 9934 and
10824 nodes respectively, the cubic and linear trend lines became very close to each other.
Fig.~\ref{fig:Scalability} in Section \ref{sec:Discussion} shows the combined testing results but includes only trees with fewer than 9000 nodes. In other words, it includes all data points in Figs.~\ref{fig:S-Data}(a,b,d).
Fig.~\ref{fig:S-Images} shows seven examples of the trees tested.

\begin{figure}[t]
    \centering
    \includegraphics[width=70mm]{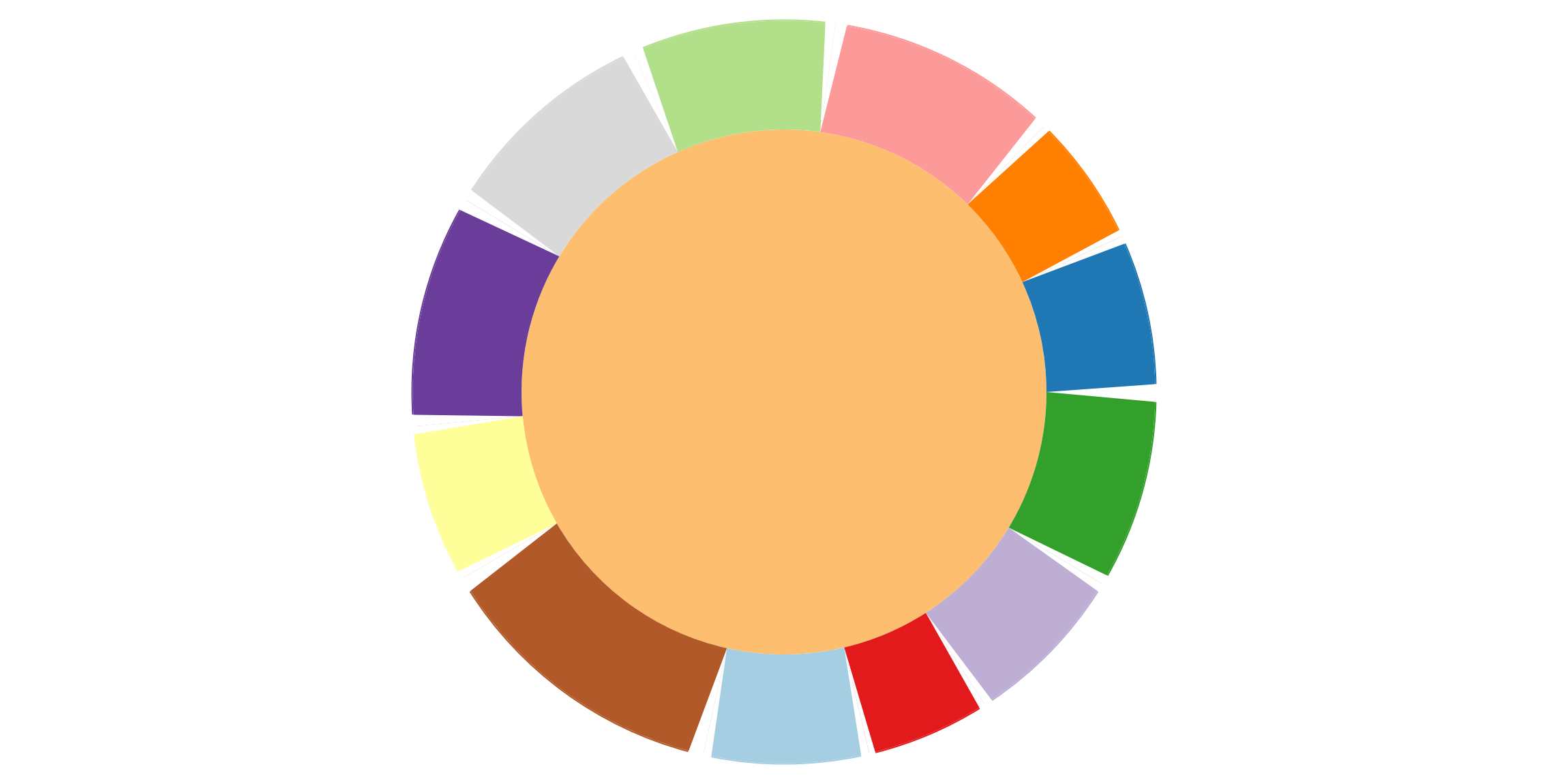}\\
    Depth = 2\\[4mm]
    \includegraphics[width=70mm]{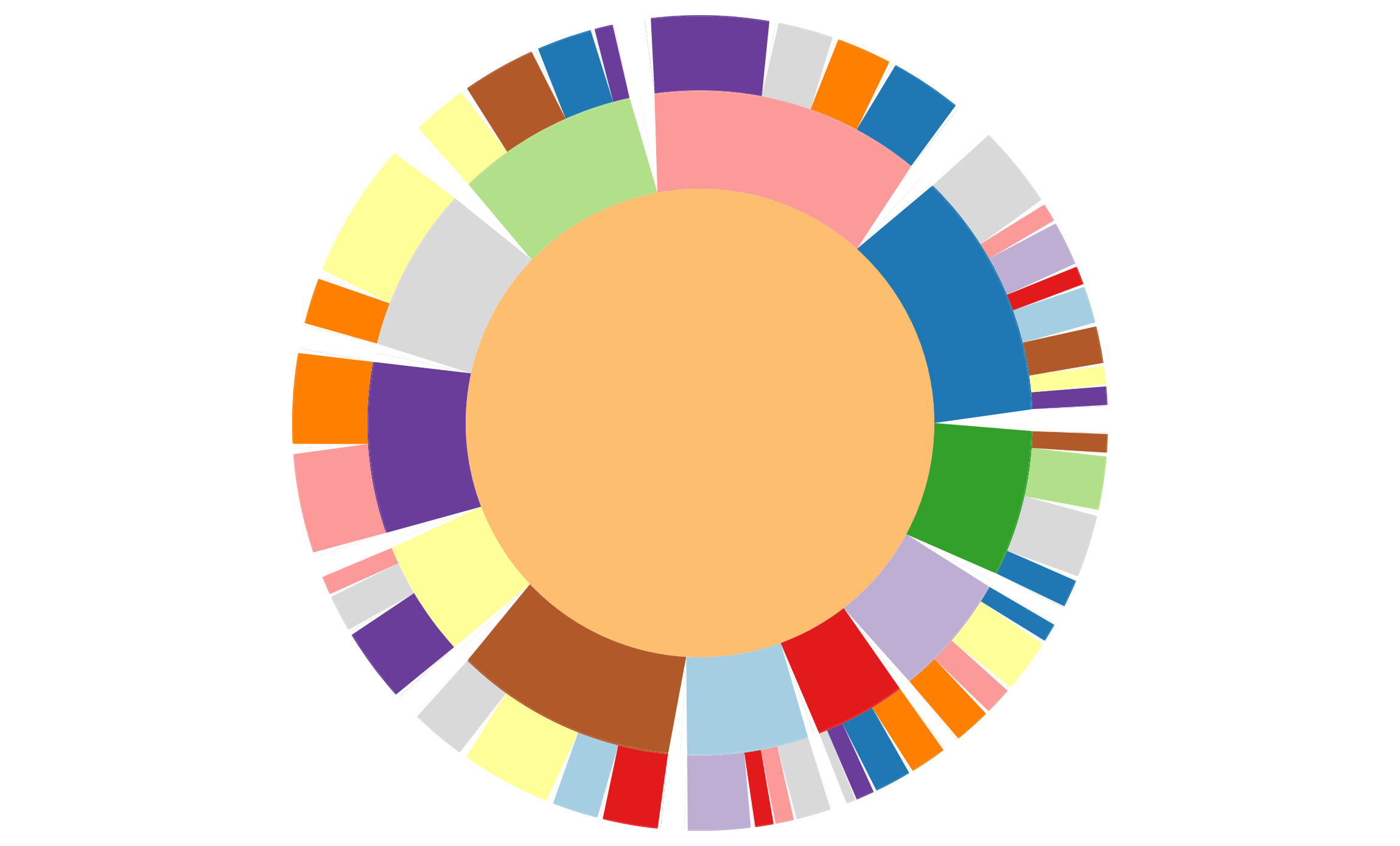}\\
    Depth = 3\\[4mm]
    \includegraphics[width=70mm]{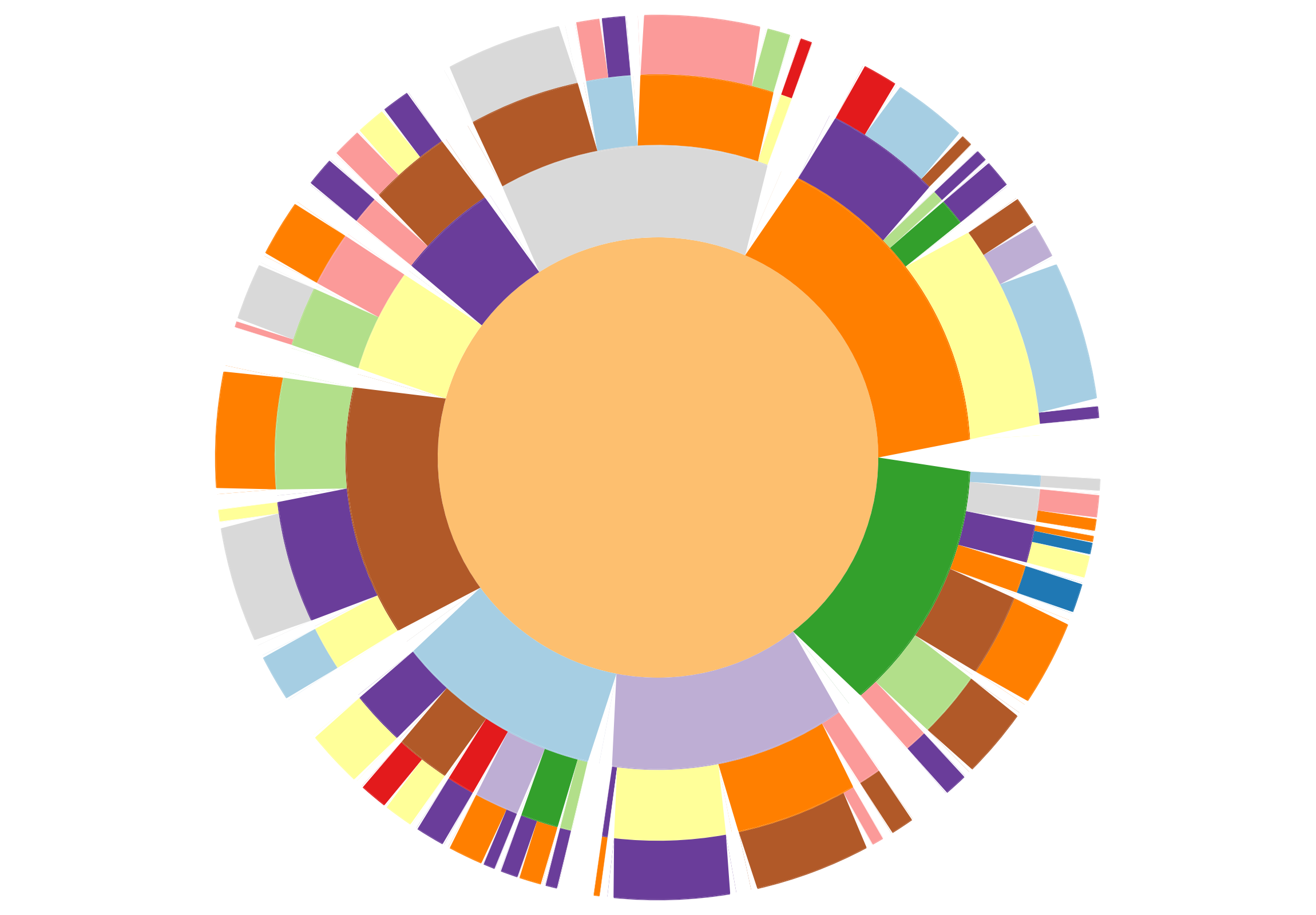}\\
    Depth = 4\\[4mm]
    \includegraphics[width=70mm]{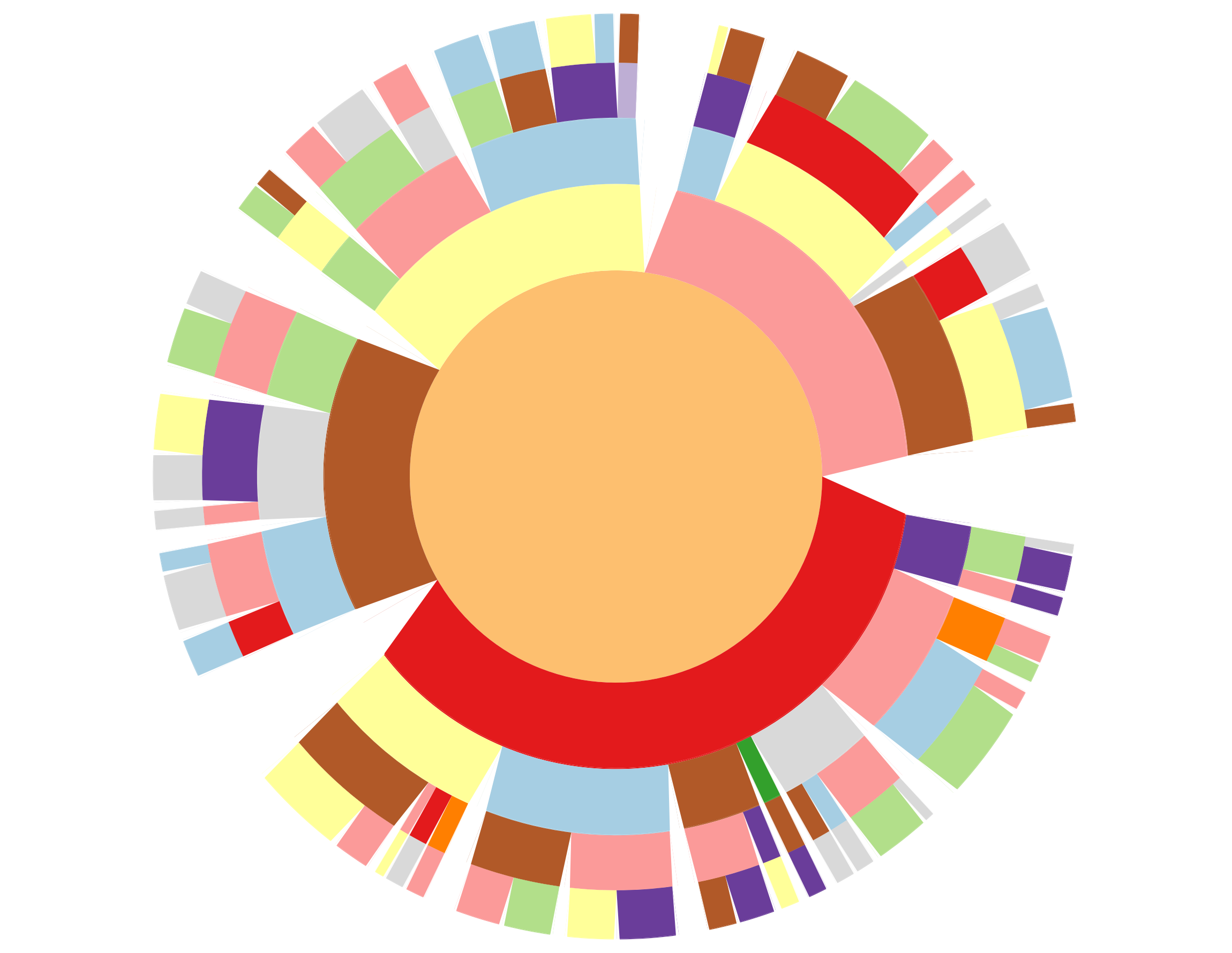}\\
    Depth = 5
    \caption{Examples of the trees that were tested.}
    \label{fig:S-Images}
    \vspace{-4mm}
\end{figure}

\begin{figure}[t]
    \centering
    \ContinuedFloat
    \includegraphics[width=70mm]{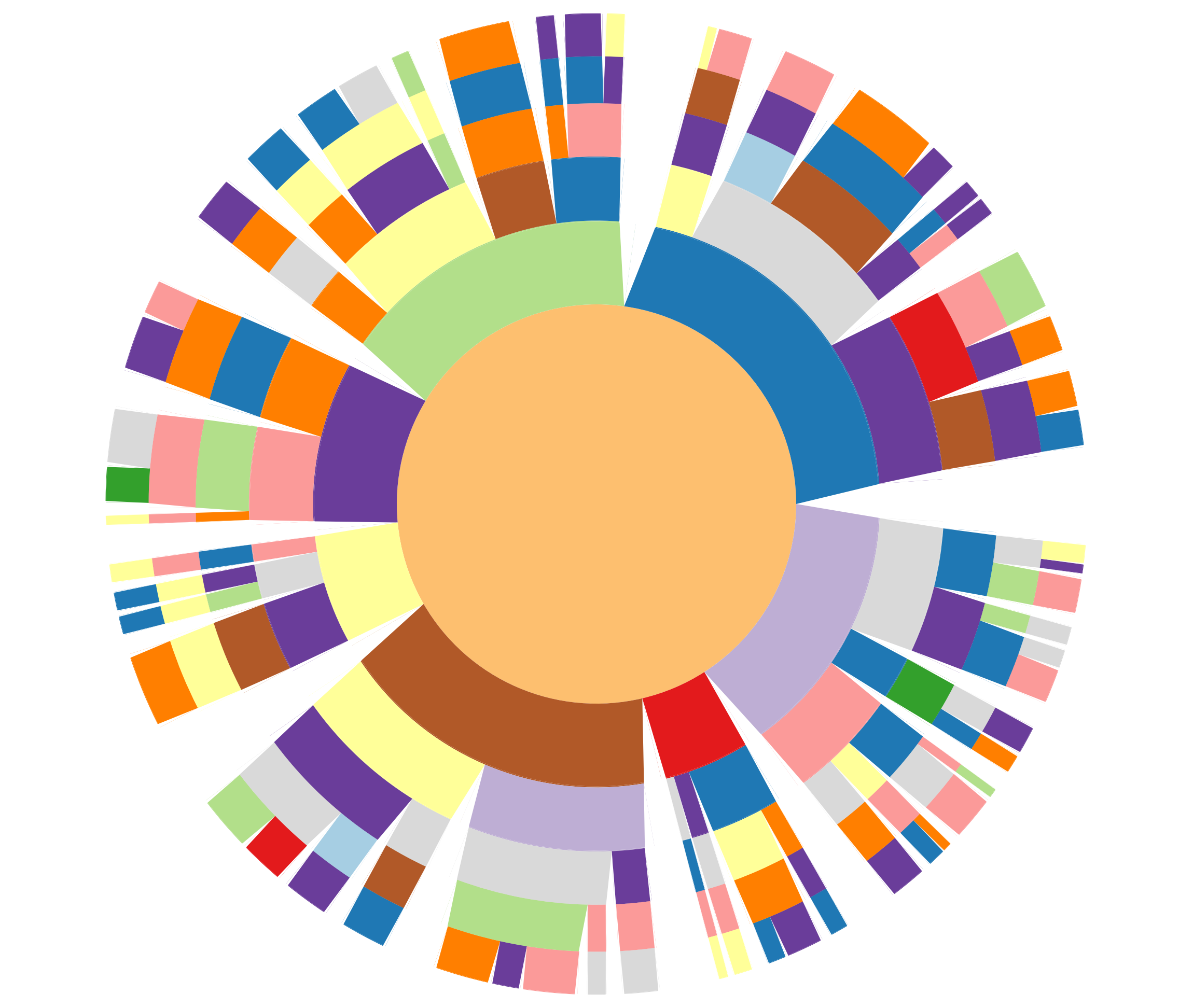}\\
    Depth = 6\\[2mm]
    \includegraphics[width=70mm]{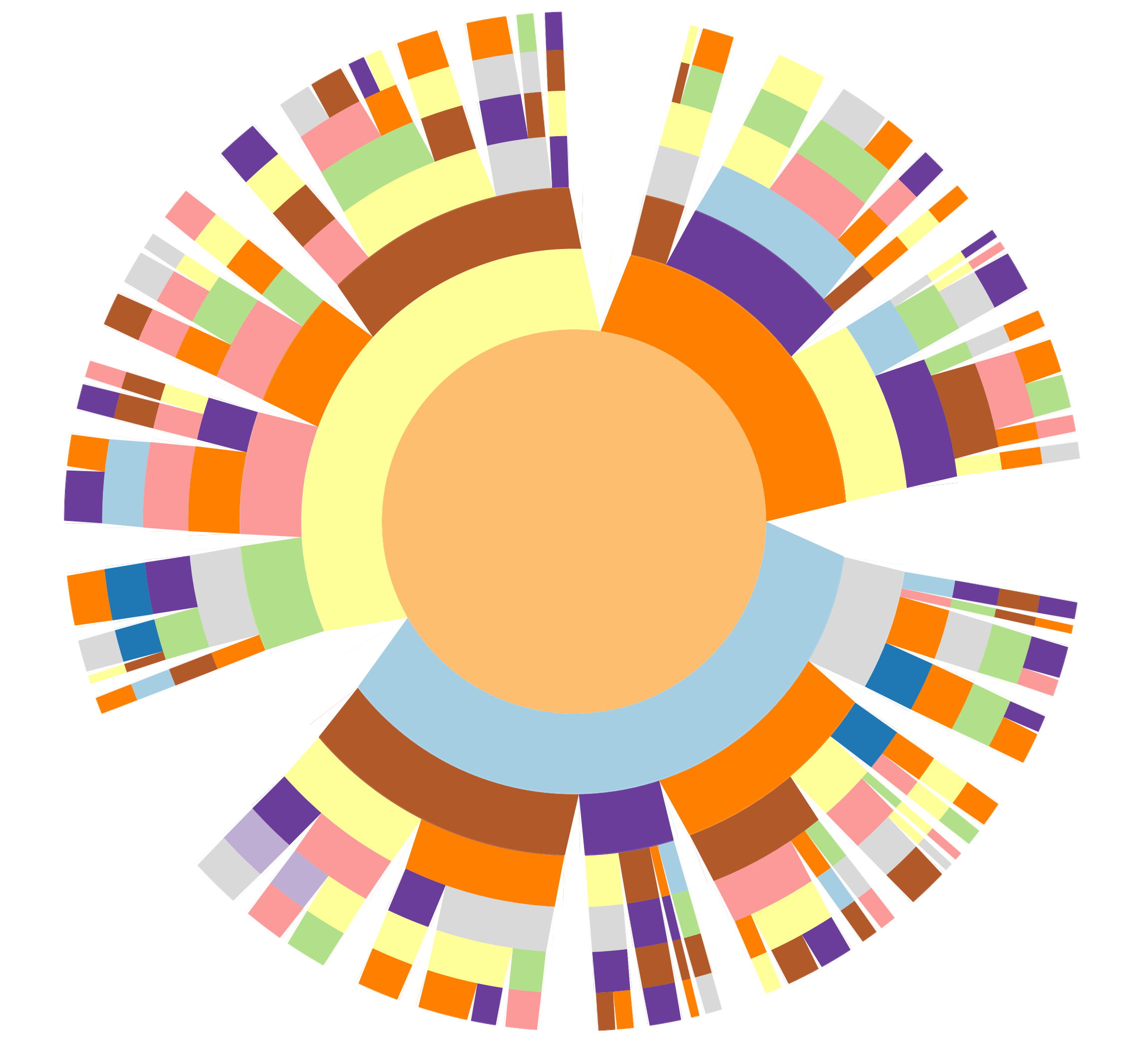}\\
    Depth = 7\\[2mm]
    \includegraphics[width=70mm]{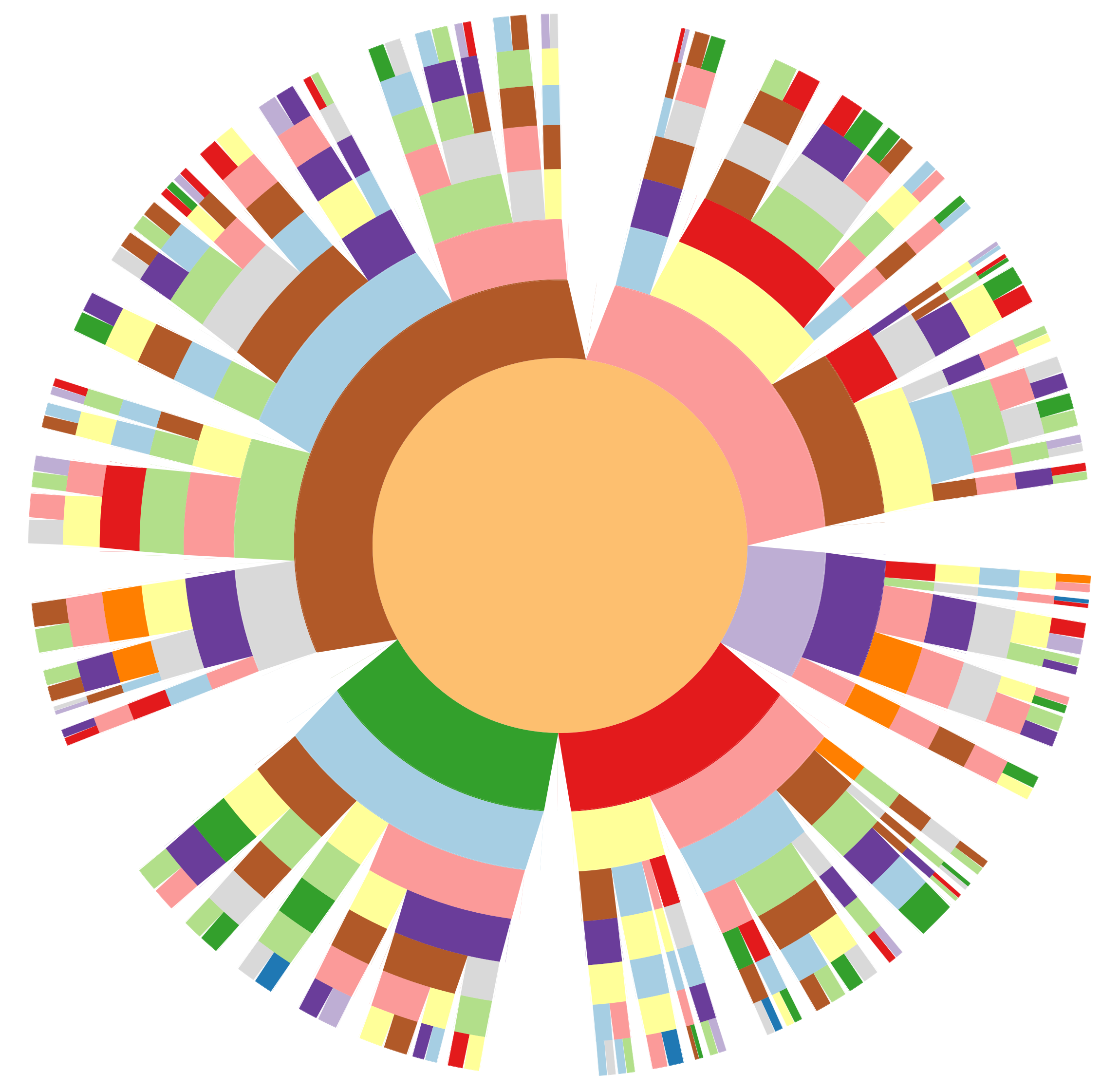}\\
    Depth = 8\\[2mm]
    \caption[]{Examples of the trees that were tested (continued).}
    \label{fig:S-Images2}
    \vspace{-4mm}
\end{figure}